\documentclass[fleqn,usenatbib]{mnras}

\usepackage{newtxtext,newtxmath}
\usepackage[T1]{fontenc}
\usepackage{ae,aecompl}
\usepackage{tabularx,booktabs,amsmath,verbatim, mathrsfs, hhline, adjustbox,gensymb,rotating}
\usepackage[version=3]{mhchem}% loads also graphicx
\usepackage{threeparttable}

\newcommand{\kms}{\mathrm{km \, s^{-1}}}
\newcommand{\lyb}{Ly\,$\beta$}

\newcommand{\zabs}{$z_{\rm abs}$}
\newcommand{\FeH}{\mathrm{[Fe/H]}}
\newcommand{\XH}{\mathrm{[X/H]}}

\newcommand{\OH}{\mathrm{[O/H]}}

\newcommand{\cmcm}{\mathrm{cm}^{-2}}

\newcommand{\keckhires}{\textit{Keck}/HIRES}
\newcommand{\hststis}{\textit{HST}/STIS}
\newcommand{\hstcos}{\textit{HST}/COS}
\newcommand{\vltuves}{\textit{VLT}/UVES}
\newcommand{\hstfos}{\textit{HST}/FOS}
\newcommand{\hstghrs}{\textit{HST}/GHRS}

\title[Low-Metallicity CGM gas at $z=0.5$]{Discovery of extremely low-metallicity circumgalactic gas at $\mathbf{z = 0.5}$ toward Q0454-220}

\author[J. M. Norris et al.]{
Jackson M. Norris,$^{1,2}$\thanks{E-mail: norris@astro-osaka.jp}
Sowgat Muzahid,$^{3}$
Jane C. Charlton,$^{1}$
Glenn G. Kacprzak,$^{4,5}$
\newauthor
Bart P. Wakker$^{6},$
and Christopher W. Churchill $^{7}$
\\
$^{1}$Department of Astronomy and Astrophysics, The Pennsylvania State University, 525 Davey Lab, University Park, State College, PA 16802, USA\\
$^{2}$Department of Earth and Space Science, Osaka University, Toyonaka, Osaka, 560-0043, Japan\\
$^{3}$IUCAA, Post Bag 04, Ganeshkhind, Pune, India, 411007\\
$^{4}$Centre for Astrophysics and Supercomputing, Swinburne University of Technology, Victoria 3122, Australia\\
$^{5}$ARC Centre of Excellence for All Sky Astrophysics in 3 Dimensions (ASTRO 3D)\\
$^{6}$Department of Astronomy, University of Wisconsin, Madison, WI 53706, USA\\
$^{7}$Department of Astronomy, New Mexico State University, Las Cruces, NM 88003, USA
}

\date{Accepted 2021 June 21. Received 2021 June 21; in original form 2020 August 27}

\pubyear{2021}

\begin{document}
\label{firstpage}
\pagerange{\pageref{firstpage}--\pageref{lastpage}}
\maketitle

\begin{abstract}
We have obtained new observations of the absorption system at $z_\mathrm{abs}=0.48$ toward QSO Q0454-220, which we use to constrain its chemical and physical conditions.
The system features metal-enriched gas and previously unknown low-metallicity gas detected $\sim 200 \, \kms$ blueward of the metal-enriched gas.
The low-metallicity gas is detected in multiple Lyman series lines but is not detected in any metal lines.
Our analysis includes low-ionization (e.g., \ion{Fe}{ii}, \ion{Mg}{ii}) metal lines, high-ionization (e.g., \ion{C}{iv}, \ion{O}{vi}, \ion{N}{v}) metal lines, and several Lyman series lines.
We use new UV spectra taken with \hstcos~ along with data taken from \hststis, \keckhires, and \vltuves.
We find that the absorption system can be explained with a photoionized low-ionization phase with $\FeH \sim -0.5$ and $n_\mathrm{H} \sim 10^{-2.3} \, \mathrm{cm}^{-3}$, a photoionized high-ionization phase with a conservative lower limit of $-3.3 < \FeH$ and $n_\mathrm{H} \sim 10^{-3.8} \, \mathrm{cm}^{-3}$, and a low-metallicity component with a conservative upper limit of $\FeH < -2.5$ that may be photoionized or collisionally ionized.
We suggest that the low-ionization phase may be due to cold-flow accretion via large-scale filamentary structure or due to recycled accretion while the high-ionization phase is the result of ancient outflowing material from a nearby galaxy.
The low-metallicity component may come from pristine accretion.
The velocity spread and disparate conditions among the absorption system's components suggest a combination of gas arising near galaxies along with gas arising from intergroup material.
\end{abstract}


\section{Introduction} \label{sec:intro}
Galactic evolution is heavily dependent on the nature of the circumgalactic material (CGM) surrounding galaxies.
Galaxies and the CGM share important accretion and feedback processes that are complex and not fully understood.
The interaction between galaxies and their surrounding material is important to understand topics such as star formation, active galactic nuclei (AGN) feedback, and galaxy evolution.
The CGM is essential for understanding cosmic metal enrichment since it holds a large percentage of the metals in the universe \citep{peeples+14} and 35\% or more of the baryons around $L \approx L^*$ galaxies \citep{werk+14}.

Numerical simulations indicate that dark matter haloes with mass $M \lesssim 10^{12} M_{\sun}$ undergo accretion of cold gas with temperature of $T \sim 10^4$\textendash$10^5$ K that interacts with outflowing gas from stellar feedback and AGNs.
The cold-accretion is fueled by large scale filamentary structure from the cosmic web and tends to form warped thick discs of material that co-rotate with the galaxy disc \citep[e.g.][]{stewart+13,ford+14,stewart+17,hafen+17,hafen+19}.
The net result of these inflows and outflows is a complex structure of material that can enrich the metallicity of both the CGM and intergalactic medium (IGM) while interacting with the interstellar medium (ISM).
Most of the metals not found in the ISM within $R \sim 0.8 R_\mathrm{vir}$ may be produced from gas processed through recycled accretion, wherein metals originating from an outflowing wind enrich the CGM and are then re-accreted onto the galaxy within a few Gyr \citep{ford+14}.

Characteristics of the CGM have been categorized in a number of studies that look at statistical samples of quasar absorption line (QAL) systems.
Absorption systems near galaxies tend to be metal-poor with metallicities between $-2 < \XH < -1$ \citep{tripp+05,cooksey+08,kacprzak+10b,ribaudo+11b,churchill+12} or metal-enriched with metallicities of $-0.7 < \XH$ \citep{thom+11,stocke+13,muzahid14,muzahid+15,hussain+15,bouche+16,rosenwasser+18}.
The probability density function of metallicity of 32 QAL systems within 150 kpc of $L^*$ galaxies at $z \sim 0.2$ from the COS-Haloes dataset is consistent with unimodality with a median $\XH = -0.51$ and a 95\% confidence interval of $\mathrm{[-1.71,0.76]}$ \citep{prochaska+17}.
Historically, one method for studying the CGM at $z<1$ involves further observations of previously-observed quasars, wherein the CGM is often detected in absorption ionization transitions detectable from ground-based observations, most notably \ion{Mg}{ii} \citep[e.g., ][]{lanzetta+1987, steidel+1992, churchill+1999};
this method can introduce a bias towards systems with higher metallicities since only systems with detectable metals are observed.
By looking at Lyman limit systems (LLSs) and partial Lyman limit systems, which tend to be selected based on the presence of hydrogen only, it is possible to partially alleviate this bias.
For LLSs at redshift $z \lesssim 1.0$, the metallicity distribution is bimodal \citep{lehner+13}, with the lower and upper modes recently refined to be $\langle\mathrm{[X/H]}\rangle = -1.76 \pm 0.09$ and $\langle\mathrm{[X/H]}\rangle = -0.33 \pm 0.077$ \citep{wotta+16}, although there is some indication that this bimodal nature only occurs within the range $0.45 \lesssim z \lesssim 1$ \cite{wotta+2019, lehner+19}.
Some of the lowest metallicity determinations in Lyman limit systems at $z \sim 0.5$ range from $-2.6 < \FeH < -2.9$ \citep{wotta+16,pointon+19, lehner+19}.
Despite these rare examples, there is a dearth of extremely low-metallicity QAL systems. This also applies to weak \ion{Mg}{ii} absorbers ($W_{r}^{2796} < 0.3 \, $\AA) at $z_\mathrm{abs} < 0.3$ \citep{muzahid+18}, suggesting that most QAL systems at lower redshifts are composed of material that has been affected by inflow/outflow processes from galaxies.

At higher redshifts, although there are examples of metal-enriched systems \citep[e.g.,][]{crighton+15}, metallicity is generally lower.
Some extreme examples include systems with
$z=3.41$, $\XH < -4.2$;
$z=3.10$, $\XH < -3.8$ \citep{fumagalli+11};
$z=3.2$, $\XH = -3.35 \pm 0.05$ \citep{fumagalli+15};
and $z=4.4$, $\XH < -4.14$ \citep{robert+19}.
More recently, the population of $z \sim 2.5 -3.5$ LLSs has been found to have a peak at $\XH \sim -2$ and the probability of a system having $\XH \lesssim -3$ is about $10\%$ \citep{fumagalli+15}.
A population of 17 LLSs from the Sloan Digital Sky Survey at $z \sim 3.2$--$4.4$ has been found to be devoid of metal lines, where 8 LLSs have metallicities between $\XH = -2$ to $-3$ and 9 LLSs have upper limits from $\XH < -1.68$ to $< -3.08$ \citep{cooper+15}.
In general, systems with metallicities below $-2.0 < \XH$ are associated with high redshift.

The QSO Q0454-220 contains an absorption system at $z=0.48382$ that produces a full Lyman limit break.
The system was originally observed for its \ion{Mg}{ii} metal-enriched component with an equivalent width of $W_\mathrm{r}(2796) = 0.426 \pm 0.006$\ \AA\ \citep{churchill+01}, but new observations with the \textit{Hubble Space Telescope}/Cosmic Origins Spectrograph (\hstcos) with coverage of \ion{O}{vi} and the Lyman series have revealed the existence of an extremely low-metallicity subsystem devoid of metal lines at $\sim 200 \, \kms$ blueward of the \ion{Mg}{ii} component that is the lowest metallicity absorption system found below $z < 1$.
The system thus simultaneously contains a metal-enriched component and an extremely metal-poor component.
In this paper, we present a model of the Q0454-220 absorption system as multiple phases of material, with each phase comprising of several clouds that each have separate fits for both $\FeH$ and ionization parameter ($U$).
It is important to disentangle the different phases and absorption components of the system because a line of sight may intersect many disparate clouds of gas that can vary greatly in their gas densities and metallicities; using the integrated \ion{H}{i} and metal column densities along a line of sight does not allow for us to observe the complex multiphase structure of the absorbing gas \citep{zahedy+2019}.

This paper is structured as follows: In \S \ref{sec:observations}, we provide details of observations of both the QAL system and neighboring galaxies.
In \S \ref{sec:method} we explain the methodology of our modeling procedure.
In \S\S \ref{sec:lowzcloud}, \ref{sec:alternative_hi}, \ref{sec:lowucloud}, and \ref{sec:highucloud}, we explain the details of the models for the low-metallicity subsystem clouds, low-ionization clouds, and high-ionization clouds, respectively.
In \S \ref{sec:caveats}, we explain some important caveats to our modeling.
In \S \ref{sec:summary}, we summarize the results of our modeling.
In \S \ref{sec:discussion}, we discuss the possible origin of the different components of the absorption system.

\section{Observations} \label{sec:observations}
\subsection{Absorption Spectra}
Spectra of QSO Q0454-220 were obtained with five instruments: W. M. Keck Observatory/High Resolution Echelle Spectrometer (\keckhires), \textit{Very Large Telescope}/Ultraviolet and Visual Echelle Spectrograph (\vltuves), \textit{Hubble Space Telescope}/Space Telescope Imaging Spectrograph (\hststis), \textit{Hubble Space Telescope}/Goddard High-Resolution Spectrograph (\hstghrs), and \hstcos.
See Table \ref{tab:observations} for a summary of the observations, which cover a wide range of chemical transitions in ultraviolet (UV) and optical spectra.
Lower resolution \textit{Hubble Space Telescope}/Faint Object Spectrograph (\hstfos) spectra (PID 1026; PI Burbidge), obtained with the G130H, G190H, and G270H gratings are also available.
The G130H spectrum shows a full Lyman limit break at $\sim 1360$~{\AA}, with $N_\mathrm{HI} > 17.74$ \citep{ribaudo+11a}.
The Lyman limit break was shown by \citet{churchill+00} to arise from this system rather than a separate $z=0.4744$ system along the same line of sight.

\begin{table*}
 \caption{Observation Information} 
 \label{tab:observations}
 \begin{tabular}{cccccc}
 \hline
  Instrument & Program & Central Wavelength & Resolution & Exposure Time & Date \\
             & ID      &     [\AA]          &            & [s]           &      \\
  \hline
  \keckhires          & ---             &  4982.4 & $\sim 45,000$         & 5400       & 1995 Jan 22\\
  \vltuves, HER-5     & 076.A-0563(A)   &  3460   & $\sim 41,400$         & 3600       & 2006 Feb 22 00:35:11\\
  \vltuves, HER-5     & 076.A-0563(A)   &  3460   & $\sim 41,400$         & 1800       & 2006 Feb 22 01:37:51\\
  \vltuves, HER-5     & 076.A-0563(A)   &  3460   & $\sim 41,400$         & 3100       & 2006 Feb 23 00:17:19\\
  \hstcos, G160M      & 12466           &  1577   & $\sim 13,000-20,000$  & 1849.344   & 2012 Apr 04 06:51:37\\
  \hstcos, G185M      & 12536           &  1850   & $\sim 16,000-20,000$  & 13865.920  & 2012 Mar 17 02:55:40\\
  \hstcos, G185M      & 12536           &  1850   & $\sim 16,000-20,000$  & 10937.504  & 2012 Mar 16 04:34:57\\
  \hststis, E230M     & 8672            &  1978   & $\sim 30,000$         & 2377.000   & 2001 Apr 27 15:23:25\\
  \hststis, E230M     & 8672            &  1978   & $\sim 30,000$         & 2906.000   & 2001 Apr 27 16:47:08\\
  \hststis, E230M     & 8672            &  1978   & $\sim 30,000$         & 2906.000   & 2001 Apr 27 18:23:29\\
  \hststis, E230M     & 8672            &  1978   & $\sim 30,000$         & 2377.000   & 2001 Apr 28 18:42:02\\
  \hststis, E230M     & 8672            &  1978   & $\sim 30,000$         & 2340.000   & 2004 Jan 14 23:30:38\\
  \hststis, E230M     & 8672            &  1978   & $\sim 30,000$         & 2400.000   & 2004 Jan 15 00:56:04\\
  \hstfos, G130H      & 1026            &  1600   & $\sim 1,300$          & 1620.000   & 1991 Apr 16 00:20:17\\
  \hstfos, G130H      & 1026            &  1600   & $\sim 1,300$          & 1620.000   & 1991 Apr 16 01:55:05\\
  \hstfos, G130H      & 1026            &  1600   & $\sim 1,300$          & 1629.973   & 1991 Apr 16 03:32:14\\
  \hstfos, G130H      & 1026            &  1600   & $\sim 1,300$          & 1629.973   & 1991 Apr 16 05:08:35\\
  \hstcos, G225M      & 13398           &  2283   & $\sim 20,000-24,000$  & 1950.432   & 2013 Dec 20 12:56:39\\
  \hstghrs, G200M     & 5961            &  2000   & $\sim 22000$          & 10009.613  & 1995 Oct 12 14:00:11\\
  \hstghrs, G200M     & 5961            &  2000   & $\sim 22000$          & 10009.613  & 1995 Oct 12 19:18:35\\
  \hstghrs, G200M     & 5961            &  2000   & $\sim 22000$          & 10009.613  & 1995 Oct 24 12:25:35\\
  \hstghrs, G200M     & 5961            &  2000   & $\sim 22000$          & 10009.613  & 1995 Oct 24 18:46:35\\
  \hstghrs, G200M     & 5961            &  2000   & $\sim 22000$          & 10009.613  & 1995 Oct 11 13:52:29\\
  \hline
 \end{tabular}
\end{table*}

The \keckhires\ spectrum covers wavelengths 3766 to 6199\AA\ at a resolution of $R = 45,000$ and was originally presented in \citet{churchill+01}.
The data were reduced with the IRAF5 APEXTRACT package (v2.10.3) for echelle data as described in \citet{churchill+01} and references therein.
This wavelength range corresponds to a rest wavelength range of 2538 to 4178~{\AA} and includes coverage of lines from the ions \ion{Mg}{ii}, \ion{Mg}{i}, \ion{Mn}{ii}, and \ion{Ca}{ii}.
Some \ion{Fe}{ii} multiplet lines are also included in the \keckhires\ coverage, but we prefer to use the \ion{Fe}{ii} lines from the \vltuves\ spectrum in our analysis since they are covered at a higher signal-to-noise ratio (SNR).

\vltuves\ observations \citep{dekker+2000} with $R \sim 40,000$ were obtained as part of program 076.A-0463(A) (PI: Sebasti\'{a}n L\'{o}pez).
The data were reduced using the European Southern Observatory (ESO)/\textit{VLT} UVES Post Pipeline Echelle Reduction (UVES\_popler) code\footnote{\url{http://astronomy.swin.edu.au/~mmurphy/UVES\_popler/}} \citep{murphy+2019}.
The QSO was observed on February 22-23, 2006 for 8500 seconds.
Although the \vltuves\ data do not cover \ion{Mg}{ii} or \ion{Mg}{i}, this spectrum provided the highest SNR of the multiple \ion{Fe}{ii} lines at 2344, 2374, 2383, 2587, and 2600~{\AA}.
Because of their range of oscillator strengths, these \ion{Fe}{ii} multiplet profiles provided the best Voigt profile fit parameters ($N_\mathrm{FeII}$ and Doppler parameter $b_\mathrm{Fe}$) to characterize the low-ionization phase.
These values are used as constraints for our Cloudy photoionization models.

\hstcos\ data were taken with the G160M grating under program 12466 and the G185M grating under program 12536 as specified in Table~\ref{tab:observations}.
The \hstcos data were first reduced with the \texttt{CALCOS} pipeline software, then the wavelength calibration was further corrected using custom software that cross-correlates important lines as described in the appendix of \citet[especially appendix]{wakker+2015}.
The alignment was based on ISM lines (including \ion{Al}{iii} and \ion{Si}{ii} transitions) and lines from other intergalactic systems at higher redshift along the same line of sight (such as Ly$\alpha$, Ly$\beta$, \ion{C}{iv}, \ion{C}{iii}, \ion{O}{i}, and others).
These UV data provide coverage of the lines \ion{C}{ii} 1036, \ion{C}{iii} 977, \ion{N}{ii} 1084, \ion{N}{iii} 990; the doublets \ion{S}{vi} 933/945, \ion{O}{vi} 1032/1038, \ion{Si}{iv} 1394/1403; and the \ion{H}{i} Lyman series lines ($\beta$ through $\epsilon$).
\ion{S}{vi} 945 is blended with the redward wing of the \ion{H}{i} 950 line from the $z=0.47441$ system along the same line of sight.
\ion{S}{vi} 933 is found close to the blue edge of the G160M grating data, and so we do not use \ion{S}{vi} 933 in our analysis; \ion{S}{vi} can provide a good diagnostic of conditions in extremely high-ionization material, so it is unfortunate that we do not have better coverage of these lines.

A few other lines are also blended with the absorption system centred at $z=0.4744$.
The \lyb\ and \ion{O}{vi} 1032 lines from the $z=0.4833$ system are blended with the two \ion{O}{vi} lines from the \zabs\ = 0.4744 system.

The \hststis\ spectrum was obtained with the E230M grating with $R=30,000$ as part of program 8672 (PI: Churchill) in order to cover the \ion{C}{iv} 1548/1551 doublet for the $z=0.4744$ and $z=0.4833$ systems.
This spectrum also provides high resolution coverage of the \ion{Si}{iv} 1394/1403 doublet, \ion{Si}{ii} 1527, and \ion{Si}{i} 1562.

The \hstghrs\ spectrum was obtained with the G200M grating with $R=22,000$ as part of program 5506 (PI: John Webb).
These data are important because they cover the $\mathrm{Ly\,}\alpha$ transition, which is essential for uncovering information about deuterium (see \S \ref{sec:deuterium}) and the intermediate-ionization phase (see \S \ref{sec:intermediateucloud}).

\begin{figure*} %Figure 1
\includegraphics[angle=0,width=0.86\textwidth]{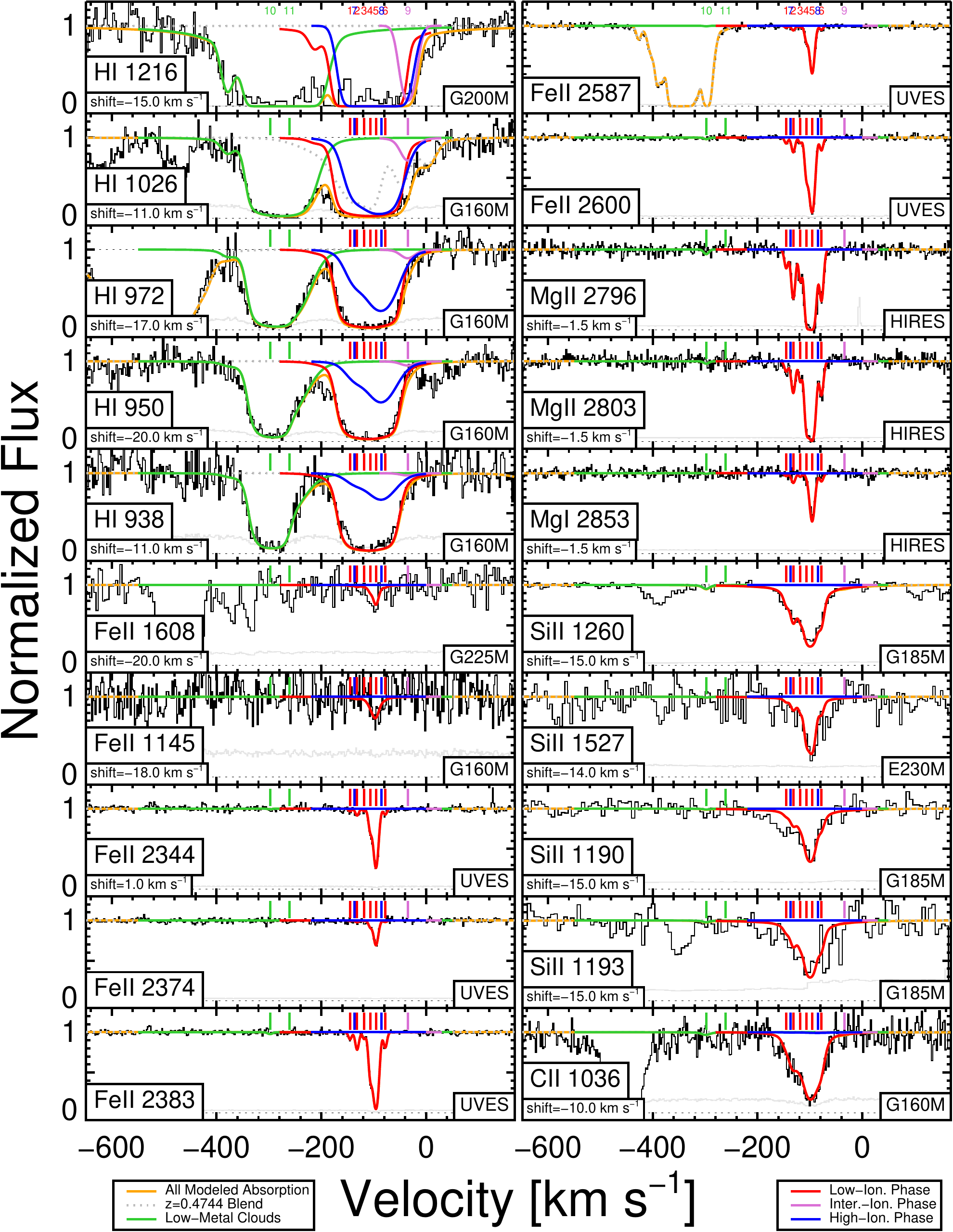}
\caption{Photoionization fit of our model to the data with the lines most associated with low-ionization.
Clouds are numbered in the top panels according to the numbers given in Table \ref{tab:cloud_parameters}.
The low-metallicity subsystem clouds are shown in green, the low-ionization clouds are shown in red, the intermediate-ionization cloud is shown in magenta, the high-ionization clouds are shown in blue, the blend with the $z=0.4744$ system along the same line of sight is shown as a dotted grey line, and all modeled absorption in aggregate is shown in orange.
The 0-point of the x-axis is set to $z=0.48382$, the redshift of the nearby G1 galaxy.
The only lines that were fit with Voigt profiles are the blueward \ion{H}{i} region and \ion{Fe}{ii} transitions.
The data are shown after applying a shift correction by the amount indicated in each panel in order to gain a good fit; this is necessary due to slight differences between the instrument or problems with the wavelength calibration.
The complex absorption feature in the \ion{Fe}{ii}2587 transition is due to a blend with the \ion{Fe}{ii}2600 transition from the $z=0.4744$ system along the same line of sight.
\label{fig:transplotq_lowu}}
\end{figure*}

\begin{figure*} %Figure 2
\includegraphics[angle=0,width=0.88\textwidth]{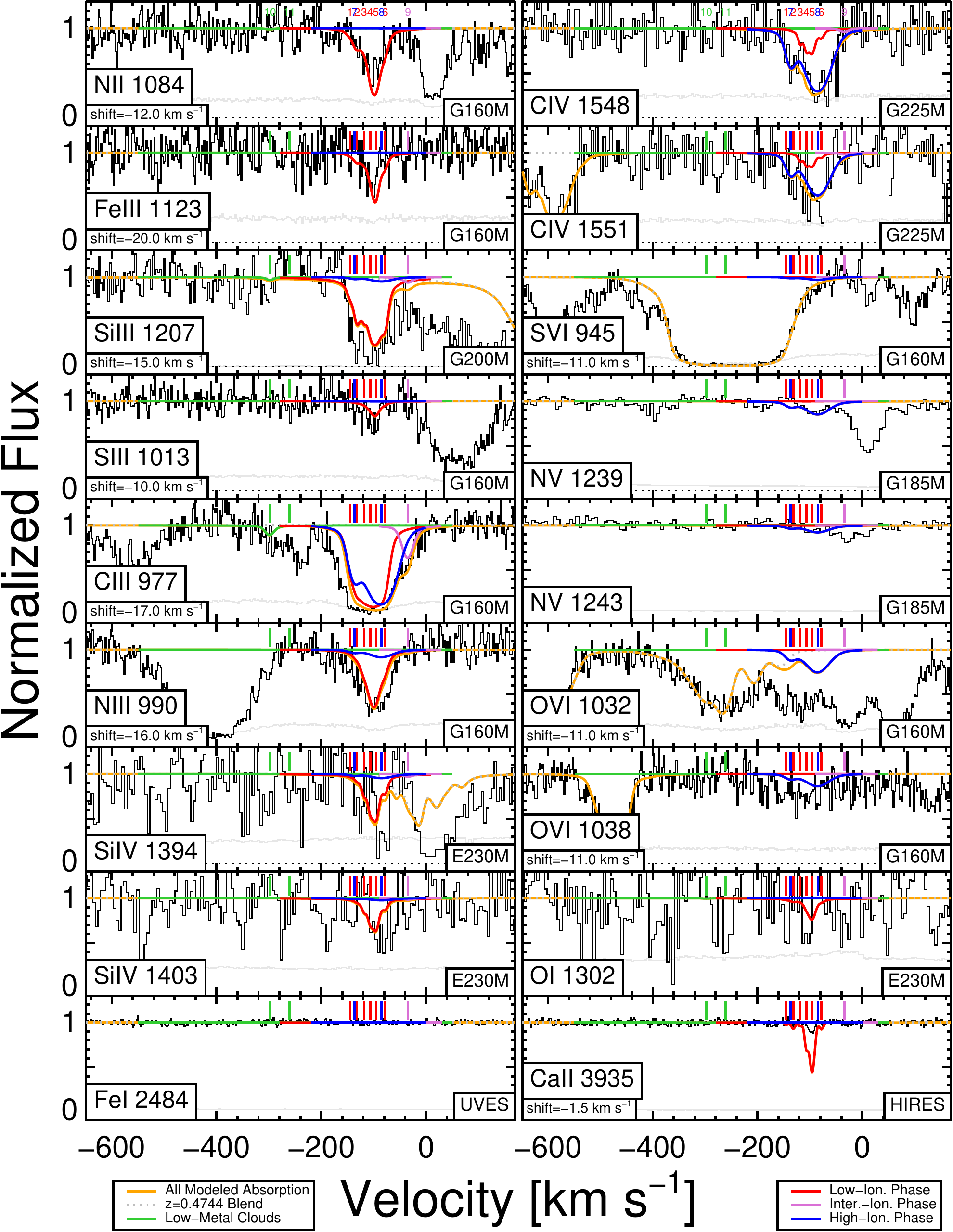}
\caption{Same as Figure \ref{fig:transplotq_lowu} but with mostly the high-ionization lines.
The only lines that were fit with Voigt profiles were the \ion{C}{iv} transitions.
The broad region in \ion{S}{vi}945 is due to a blend with the Lyman series from the $z=0.4744$ system along the same line of sight.
Similarly, some of the absorption found near the \ion{O}{vi}1032 transition is from the \ion{O}{vi}1038 transition from the $z=0.4744$ system.
\label{fig:transplotq_highu}}
\end{figure*}

The observed absorption profiles for the key constraining transitions for the $z=0.4833$ system are presented in Figure \ref{fig:transplotq_lowu} and Figure \ref{fig:transplotq_highu}.
The zero-point in velocity for the plot is set at the systemic velocity of the corresponding G1 galaxy ($z = 0.48382$), as described in \S \ref{sec:galaxy}.
The strong \ion{Mg}{ii} absorption system at $v \sim -110 \, \kms$ is detected in the Lyman series lines and in various other low-, intermediate-, and high-ionization transitions.
The absorption seen in the Lyman series at $v \sim -300 \, \kms$ does not have any corresponding metal lines detected.

In Figures \ref{fig:transplotq_lowu} and \ref{fig:transplotq_highu}, some of the lines contain velocity shifts indicating that we added those values to the velocity of the data.
This is necessary since our data were taken from multiple telescopes and there exists potential slight wavelength calibration errors.
The biggest example of an imposed velocity shift of the data is in the Lyman series, in which running a Voigt profile fit on the unshifted data does not yield satisfactory results.
This indicate something may be problematic in the data reduction.
However, since all the Lyman lines are strong within the low-metallicity subsystem, we are able to use them as a guide for a velocity correction to get better results.
The velocity shifts are always less than the resolution element of the respective spectrograph.
Every line that contains a velocity shift has its velocity shift value explicitly labelled within the figures.

\subsection{Galaxy Data} \label{sec:galaxy}
There exists a galaxy (G1) associated with the absorption system that was first spectroscopically identified by \citet{chen+98}.
Its properties were later refined by several authors \citep{kacprzak+10a,nielsen+13a,kacprzak+15}.
The galaxy's most recently determined properties are listed in Table \ref{tab:galaxy_properties}.

\begin{table}
\centering
 \caption{Galaxy Properties}
 \label{tab:galaxy_properties}
 \begin{tabular}{cl}
  \hline
  Property & Value\\
  \hline
  $z$                         & 0.48382 \\ \\
  D                         & $108.0 \pm 0.06$ kpc  \\ \\
  $\phi$                    & $85.2$\degree$^{\, +4.4}_{\, -3.7}$ \\ \\
  $i$                       & $42.1$\degree$^{\, +2.7}_{\, -3.1}$\\ \\
  $\mathrm{B}-\mathrm{K}$   & 1.78 \\ 
  \hline
 \end{tabular}
\end{table}

To summarize, G1 is a spiral galaxy at $z=0.48382$ with one long spiral arm, a compact bulge, and a perturbed morphology.
The galaxy was modeled using GIM2D \citep[Galaxy IMage 2D,][]{simard+02}, which fits the galaxy with a two-component model of a S\'{e}rsic bulge and an exponential disc.
The GIM2D modeling indicates that the absorption system lies at  an azimuthal angle $\Phi = 85.2\degree \, _{-3.7}^{+4.4}$ (near the minor axis of the galaxy) and has an inclination of $i = 42.1\degree \, _{-3.1}^{+2.7}$ ($i=0\degree$ corresponds to a face-on orientation for the galaxy).
For more information on the GIM2D modeling, refer to \citet{kacprzak+15}.
The galaxy has an absolute $B$-band magnitude of $M_B = -21.0$, which translates to $L_B = 0.90 L_B^{\ast}$, and a reddish color of $B-K = 1.78$.
The absorption system is at an impact parameter of $D = 108.0 \pm 0.6$ kpc from G1.

\citet{kacprzak+10a} use [\ion{O}{\sc II}], H$\beta$, [\ion{O}{\sc III}], H$\alpha$, and [\ion{N}{ii}] to establish a rotation curve for G1 (see Figure 3d of their paper).
They find that the projected rotation curve flattens at a maximum of $138 \, \kms$ and the mean absorption redshift is offset by $-98\ \, \kms$ from the galaxy.
The blueward side of the projected rotation curve is consistent with the velocity range of the detected metal lines (see \S\S \ref{sec:lowucloud} and \ref{sec:highucloud}), albeit at a large impact parameter implying the absorption could arise in an extended gaseous disc.
However, the low-metallicity component at $\sim -300 \, \kms$ is not consistent with an extended rotating disc model.

\begin{figure} %Figure 3
\includegraphics[angle=0,width=\columnwidth]{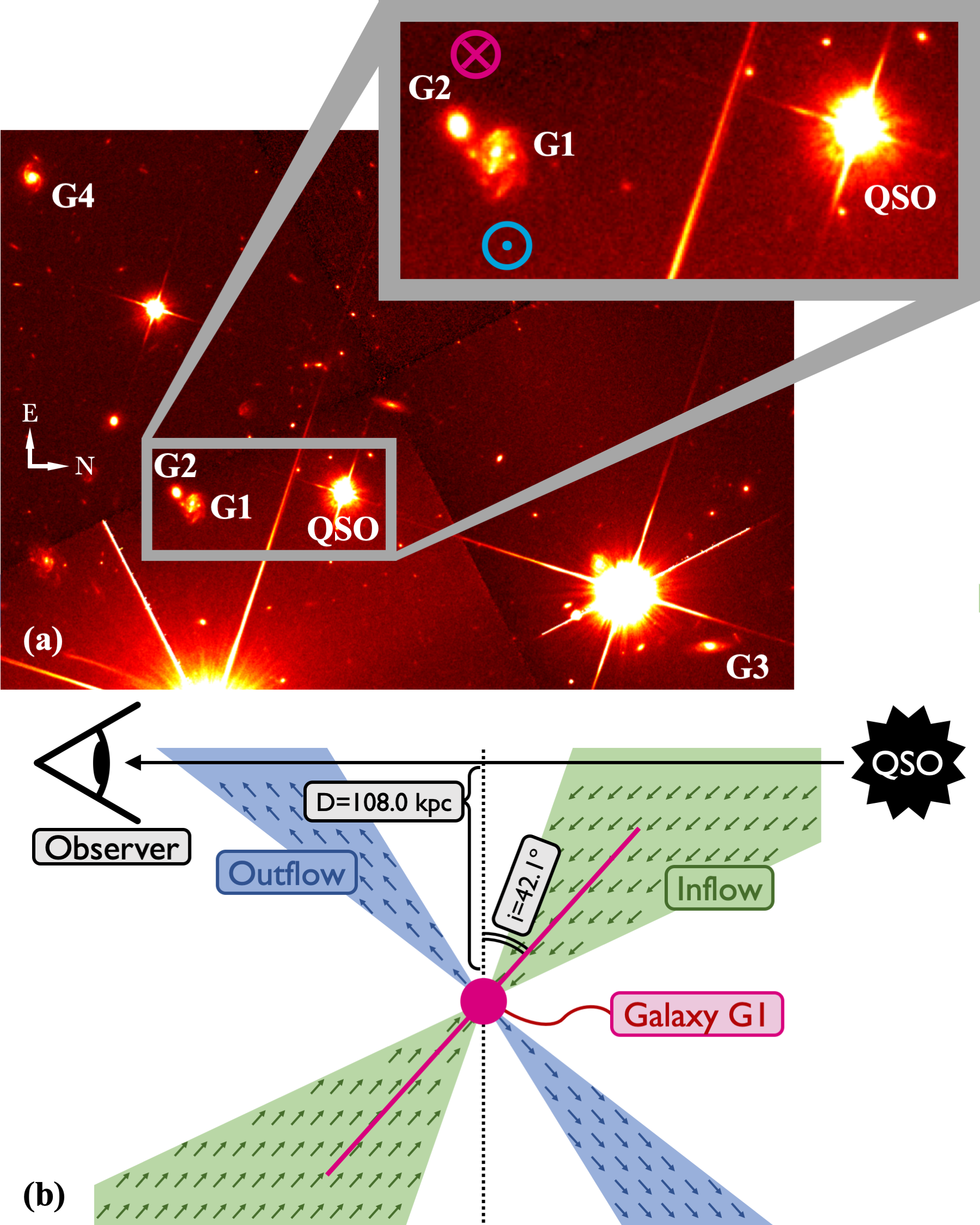}
\caption{\textbf{(a)} Image of the quasar field Q0454-220 obtained with the \textit{F702W} filter on the \textit{Hubble Space Telescope}/Wide Field and Planetary Camera 2 (\textit{HST}/WFPC2).
G1 is a spiral galaxy associated with the absorption system at $z = 0.48382$ at an impact parameter $\mathrm{D} = 108.0 \pm 0.06$ kpc.
Three other galaxies in the field are also labelled (see \S \ref{sec:galaxy}).
We derive the 3D disk orientation following the methods of \citet{Ho+20}, indicated by the magenta and blue symbols, using the direction of spiral arms and rotation curve of G1 (see \S \ref{sec:discussiongal}).  
\textbf{(b)} Using the 3D galaxy orientation, we present a simple diagram of the absorption system, G1, QSO, and observer.
Note that this diagram places the observer on the left and the QSO on the right, so the line-of-sight from the observer to the QSO goes horizontally across the diagram; in effect, the reader is seeing the system `from the side' rather than what is shown in Figure \ref{fig:diagram}(a).
Both the inflowing material along the galaxy's major axis as well as the outflowing material along the galaxy's minor axis are blueshifted.
The low- and high-ionization phases are expected to arise from a cold-flow disc, which arises due to filamentary structure (the filamentary structure is not shown in this diagram).
The low-metallicity subsystem is not aligned with a cold-flow thick disc, and is likely inflowing material at higher angular momentum.
The diagram is not to scale.
\label{fig:diagram}}
\end{figure}

There exist two other galaxies at similar redshifts to G1 in the field, but at larger impact parameters.
Galaxy G3 has $z=0.4836$ ($\Delta v = -44 \, \kms$) and $D=284.2 \, \mathrm{kpc}$.
Galaxy G4 has $z=0.4837$ ($\Delta v = -24 \, \kms$) and $D=314.7\ \mathrm{kpc}$.
The G1, G3, and G4 galaxies suggest a group environment.
There also exists a nearby foreground galaxy G2 at redshift $z=0.3818$.
See Figure \ref{fig:diagram}(a) for an image of these galaxies in the quasar field.

\section{Methodology} \label{sec:method}
Our goal is to model the absorption system with a minimum number of Voigt profile (VP) components in different ionization phases.
Henceforth, the separate VP components may be referred to as ``clouds.''
We start by choosing an ionization species with unsaturated transitions and large SNR for each ionization phase.
These ``optimized ions'' or \ion{Fe}{ii]} in the low-ionization phase (\S \ref{sec:lowucloud}), \ion{C}{iii} in the intermediate-ionization phase(\S \ref{sec:intermediateucloud}), and \ion{C}{iv} in the high-ionization phase (\S \ref{sec:highucloud}).
We use the VPFIT \citep{vpfit14} program to simultaneously fit several Voigt profiles to all transitions within a single ionization species.
Each cloud's VP fit is defined by three parameters: column density ($N_\mathrm{X}$), Doppler parameter ($b_\mathrm{X}$), and redshift ($z$).
We use the VPFIT solutions of the initial ionization species as a mathematical starting point for photoionization models and tune the model parameters so that the model is consistent with the profiles of all other ionization species.

We investigate models of photoionization equilibrium (PIE), collisional ionization equilibrium (CIE), and non-equilibrium collisional ionization (non-CIE).
For CIE and non-CIE, we adopt models of \citet{gnat+07} (G\&S), which are discussed in more detail in \S \ref{sec:cie}.
For photoionization modeling, we treat the clouds as plane-parallel slabs using version 17.00 of Cloudy \citep{ferland+13,ferland+17}.
Since the nearest galaxy to the absorption system is at a large distance, we assume the ionizing extragalactic background radiation (EBR) from \citet{haardt+12} (HM12) is the only source of ionization.
We adopt the redshift near the centre of the \ion{Mg}{ii} lines, ($z_\mathrm{EBR} = 0.48334$) to define the amplitude and shape of the EBR.
We assume solar abundance patterns for the elements based on \citet{grevesse+10}.
We find that solar abundance patterns are sufficient for us to model the data with the exception of \ion{Ca}{ii} and \ion{N}{v}, discussed in more detail in \S\ref{sec:caveats} and \S\ref{discussion:nitrogen}.
This method is similar to those described in previous studies \citep{charlton+03,ding+05}.

For each cloud, we run a grid of Cloudy models that produce the column density of the optimized ion, $N_\mathrm{X}$, determined from the VP fit.
The grid is defined by two parameters: metallicity relative to solar ($\FeH \equiv \log \frac{Z}{Z_{\sun}}$) and ionization parameter ($U \equiv \frac{n_\gamma}{n_\mathrm{H}}$, where $n_\gamma$ is the volume density of ionizing photons).
We construct grids of $\{ \log U, \FeH \}$ using our values of $z_\mathrm{EBR}$ and $N_\mathrm{X}$ order to determine which values of $\{\log U, \FeH\}$ give adequate fits to all the other ionization species that were not VP fit.
For each point in the grid, $N_\mathrm{HI}$ is changed iteratively until $N_\mathrm{X}$ is produced.

As an example of our method, we will provide a detailed summary of the process for a single cloud.
The VP fit for Cloud \#6 (see Table \ref{tab:cloud_parameters}) gives a cloud with $N_\mathrm{FeII} = 11.98$, so we construct $\{\log U, \FeH\}$ grids in which $N_\mathrm{FeII}$ is held constant but $N_\mathrm{HI}$ is allowed to vary.
For each value of $\log U$ and $\FeH$, we compare the column densities determined by the Cloudy model to the observed data across multiple transitions.
Eventually, we find a set of solutions that satisfies the $N_\mathrm{FeII} = 11.98$ criterion and also do not conflict with the shapes of the other ionization species.

\begin{table*}
\caption{Details of Cloudy Fits}  \label{tab:cloud_parameters}
   \resizebox{\textwidth}{!}{\begin{tabular}{|lclcccccc | r@{ }r@{ }r | r@{ }r@{ }r | r@{ }r@{ }r| }
    \toprule

\multicolumn{5}{|c|}{\textbf{Voigt Profile Fit}}   &  \multicolumn{13}{|c|}{\textbf{Cloudy Output}}\\     
\midrule
\multicolumn{1}{|c}{1} & \multicolumn{1}{c}{2} & \multicolumn{1}{c}{3} & \multicolumn{1}{c}{4}        & \multicolumn{1}{c}{5}  & \multicolumn{1}{c}{6} & \multicolumn{1}{c}{7}                      &  \multicolumn{1}{c}{8} & \multicolumn{1}{c|}{9}   & \multicolumn{3}{c|}{10} & \multicolumn{3}{c|}{11} & \multicolumn{3}{c|}{12}     \\     
\multicolumn{1}{|c}{Cloud} & \multicolumn{1}{c}{Opt.} & \multicolumn{1}{c}{Redshift} & \multicolumn{1}{c}{$b_\mathrm{X}$}        & \multicolumn{1}{c}{$\log N_\mathrm{X}$}  & \multicolumn{1}{c}{$\log N_\mathrm{HI}$} & \multicolumn{1}{c}{$\log N_\mathrm{H}$}                      &  \multicolumn{1}{c}{Best Temp.} & \multicolumn{1}{c|}{Best $\log$ Dens.}   & \multicolumn{3}{c|}{$\FeH$} & \multicolumn{3}{c|}{$\log U$} & \multicolumn{3}{c|}{$\log$ Size [kpc]}     \\
                           & \multicolumn{1}{c}{Ion}       &                              & \multicolumn{1}{c}{[$\kms$]} & \multicolumn{1}{c}{[$\mathrm{cm}^{-2}$]} & \multicolumn{1}{c}{[$\mathrm{cm}^{-2}$]} & \multicolumn{1}{c}{[$\mathrm{cm}^{-2}$]} &  \multicolumn{1}{c}{[K]}        & \multicolumn{1}{c|}{[$\mathrm{cm}^{-3}$]} &   Min. & Best & Max.        & Min.& Best & Max. & Min. & Best & Max.                                     \\ 
    \midrule
~1     & \ion{Fe}{ii} & 0.483101 & $ 2.0 \pm  1.8$   & $11.59 \pm 0.05$ & 16.13 & 17.52 & $1.2 \times 10^4$ & -1.8      & -1.2   & -0.8 & \ldots    & \ldots & -3.9 & -3.5      &  \ldots  & -2.1 & +0.2    \\
~2     & \ion{Fe}{ii} & 0.483168 & $ 2.9 \pm  0.6$   & $12.06 \pm 0.02$ & 17.18 & 18.79 & $1.3 \times 10^4$ & -2.0      & -1.8   & -1.3 & -0.8      & -3.9   & -3.7 & -3.7      &  -1.7    & -0.7 & -0.3   \\
~3     & \ion{Fe}{ii} & 0.483234 & $ 1.1 \pm  0.1$   & $11.54 \pm 0.07$ & 16.09 & 18.18 & $1.3 \times 10^4$ & -2.5      & -3.0   & -0.5 & \ldots    & -3.6   & -3.2 & -2.3      &  \ldots  & -0.8 & +3.0   \\
~4     & \ion{Fe}{ii} & 0.483291 & $ 2.8 \pm  0.3$   & $12.60 \pm 0.02$ & 15.77 & 17.36 & $1.2 \times 10^3$ & -2.7      & -0.1   & +0.8 & \ldots    & -3.1   & -3.0 & -2.9      &  \ldots  & -1.4 & +0.7   \\
~5     & \ion{Fe}{ii} & 0.483346 & $ 4.8 \pm  0.1$   & $13.36 \pm 0.01$ & 17.33 & 19.04 & $1.0 \times 10^4$ & -2.2      & -1.8   & -0.3 & -0.1      & \ldots & -3.5 & -2.5      &  \ldots  & -0.2 & +2.5   \\
~6     & \ion{Fe}{ii} & 0.483434 & $ 2.6 \pm  0.6$   & $11.98 \pm 0.02$ & 16.59 & 18.51 & $1.3 \times 10^4$ & -2.3      & -1.4   & -0.7 & \ldots    & -3.7   & -3.4 & -3.2      &  \ldots  & -0.7 & +0.6   \\
~7     & \ion{C}{iv}  & 0.483144 & $11.2 \pm  3.9$   & $13.30 \pm 0.13$ & 14.46 & 17.95 & $2.1 \times 10^4$ &  -3.8     & -3.3   & -0.5 & \ldots    & -2.9   & -1.9 & -1.4      &  \ldots  & +0.3 & +3.0   \\
~8     & \ion{C}{iv}  & 0.483400 & $26.6 \pm  3.2$   & $14.00 \pm 0.05$ & 15.16 & 18.65 & $2.1 \times 10^4$ &  -3.8     & -2.6   & -0.5 & \ldots    & -2.9   & -1.9 & -1.5      &  \ldots  & +1.0 & +3.0   \\ 
~9     & \ion{C}{iii} & 0.483649 & $ 9.9 \pm  3.5$   & $12.92 \pm 0.06$ & 13.80 & 16.31 & $8.1 \times 10^3$ & -3.1      & -0.1   & +0.3 & \ldots    & -2.8   & -2.6 & -2.1      &  \ldots  & -2.1 & -1.2 \\
10     & \ion{H}{i}   & 0.482347 & $15.1 \pm  2.2$   & $18.04 \pm 0.88$ & 18.04 & 19.20 & $1.3 \times 10^4$ & \ldots    & \ldots & -3.0 & -2.5      & \ldots & -3.8 & \ldots    &  \ldots  & -0.4 & \ldots \\
11     & \ion{H}{i}   & 0.482529 & $37.4 \pm 10.1$   & $15.29 \pm 0.30$ & 15.30 & 16.80 & $1.3 \times 10^4$ & \ldots    & \ldots & -3.0 & -2.5      & \ldots & -3.8 & \ldots    &  \ldots  & -2.8 & \ldots \\ 
    \bottomrule
  \end{tabular}}
  \begin{tablenotes}
  \item Parameters for each cloud used in the final fit.
  Column \#2 is the ionization species used to determine the initial VP fits, while columns \#3, \#4, and \#5 are the parameters from the VP fits.
  Columns \#6, \#7, \#8, and \#9 are derived from the 'best' Cloudy models for $\FeH$ and $\log U$, which can be found in the centers of columns \#10 and \#11.
  The left and right sides of columns \#10, \#11, and \#12 show upper and lower limits (\textit{not} $1 \sigma$ errors).
  The `best' Cloudy models for $\FeH$ and $U$ are the values we used to construct Figures \ref{fig:transplotq_lowu} and \ref{fig:transplotq_highu}; they do not correspond to a value that is statistically better than other values.
  We represent the lock of a constraint with an ellipsis.
  We do not present the fit to the grey cloud in Figures \ref{fig:transplotq_lowu} and \ref{fig:transplotq_highu} since it is from a known blend with a lower redshift system, and we do not bother to fully model the cloud.
  The VP fit errors provided for $\log N_\mathrm{X}$ and $b_\mathrm{X}$ are what is reported by the VPFIT program, but we suspect they are overestimates rather than true $1 \sigma$ error (see \S \ref{sec:caveats}).
\end{tablenotes}
\end{table*}

\begin{figure*} %Figure 4
\includegraphics[angle = 0, scale = 0.70]{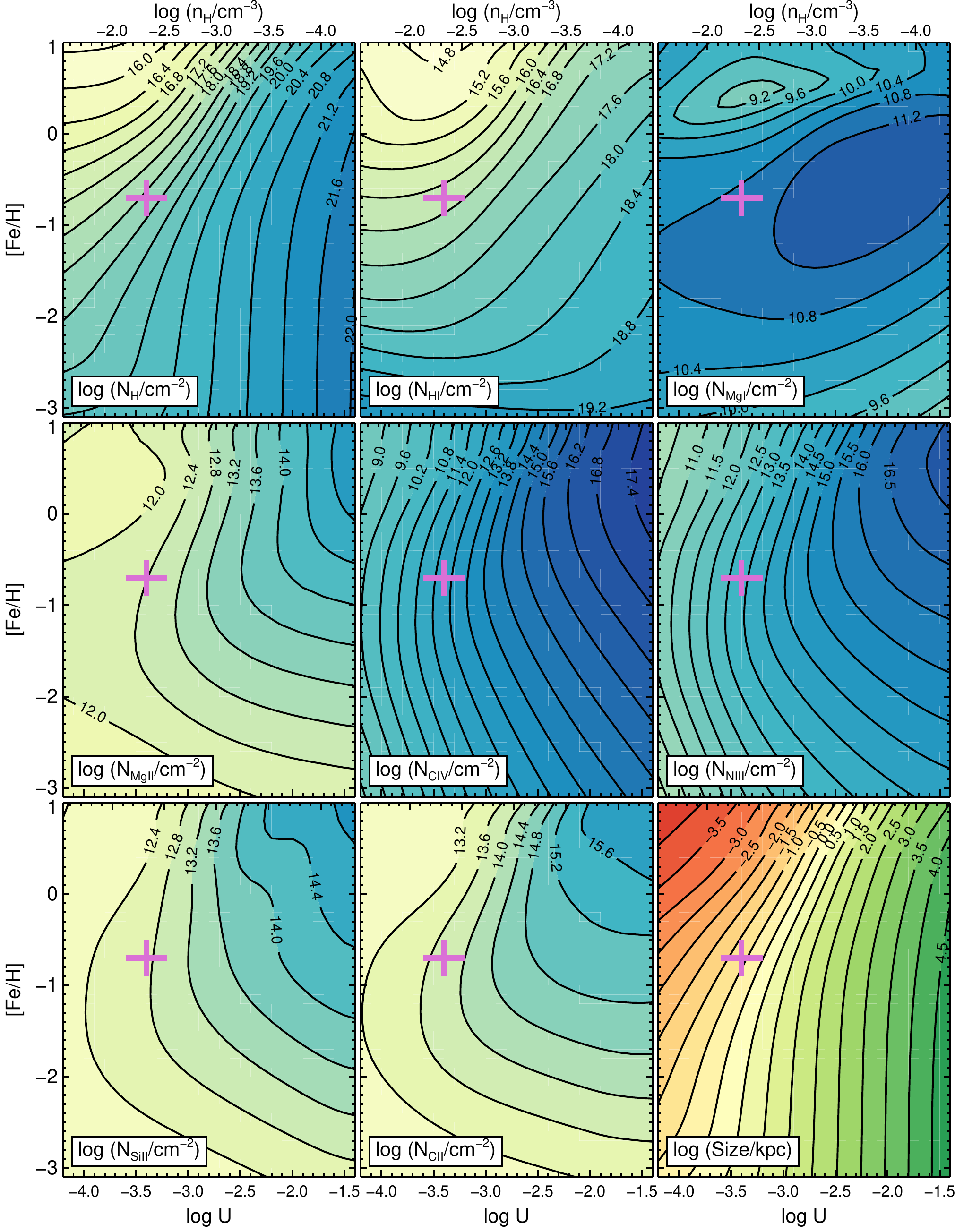}
\caption{Contour plot showing column density for several ionization species as well as size.
This figure is based on Cloud \#6 with $N_\mathrm{FeII} = 11.98$.
The cross is positioned at our `best' fit value.
Note that \ion{Mg}{i} is quite complex compared to the other species, which can make it difficult to pin down a simple range for solutions of $U$ or $\FeH$.
The color scheme for each ionization species is different, and one should not compare colors across separate ionization species (it only exists to aid the eye).
Each upper y-axis corresponds to volumetric density while each lower y-axis corresponds to ionization parameter.
\label{fig:uz_visual2}}
\end{figure*}

For certain parts of parameter space, limited constraints can be placed on metallicity based only on the metal line transitions, without reference to \ion{H}{i}.
This is useful because although the Lyman series can be used to establish metallicity limits for the clouds near the wings of the lines, clouds that appear within the central region of the saturated Lyman transitions do not have reliable \ion{H}{i} column densities.
In such cases, we are able to use fits to several metal lines concurrently to derive constraints on metallicity.
When we use metal lines, we must be careful in our modeling because solutions for metallicity at a specific value of $U$ may be degenerate.
As an example, Figure \ref{fig:uz_visual2} shows a contour plot of column densities for \ion{Mg}{i}, \ion{Mg}{ii}, \ion{H}{i}, and \ion{C}{iv} for Cloud \#6 where $\log [ N_\mathrm{FeII} / \mathrm{cm}^{-2} ] = 11.98$.
For a given value of $\log U$ there are two different values of metallicity for which $N_\mathrm{MgII} = 12.3$.
The lower metallicity solution is optically thick, while the higher metallicity solution is optically thin.
For example, if $\log U = -3.5$, then both $\FeH \approx -1.9$ and $\FeH \approx -0.7$ will provide satisfactory fits to \ion{Mg}{ii}, but intermediate values of $\FeH$ between these numbers will \textit{not} provide a satisfactory fit, producing a larger value of $N_\mathrm{MgII}$.
In this way, if \ion{Mg}{ii} were known to be $N_\mathrm{MgII} < 12.3$, then \ion{Fe}{ii} and \ion{Mg}{ii} would constrain the metallicity to be $-1.9 \nless \FeH \nless -0.7$ at this value of $U$, without the use of \ion{H}{i}.
This manner of determining limited constraints based only on metal transitions is most pronounced for \ion{Mg}{i}, where the contours create closed loops, probably due to the complex charge exchange between magnesium and hydrogen.
Complicated situations like this are not always intuitive.
Nevertheless, in \S\S \ref{sec:lowzcloud}, \ref{sec:lowucloud}, and \ref{sec:highucloud} we attempt to simplify the full space of solutions to easily specified ranges of $\FeH$ and $U$.

The Doppler parameters found from VPFIT correspond to the square root of the sum of the squares of the turbulent Doppler component, $b_\mathrm{turb}$, and the thermal component, $b_T$.
In order to determine how the Doppler parameter changes for each ionization species, we assume $b_\mathrm{turb}$ is the same across all ionization species, while $b_T = \sqrt{\frac{2 k T}{m_\mathrm{X}}}$, where $k$ is the Boltzmann constant, $m_\mathrm{X}$ is the atomic mass, and $T$ is the temperature of the gas as determined by Cloudy.
Thus, we know the column density and Doppler parameters ($N_\mathrm{X}$, $b_\mathrm{X}$) for each transition.
This allows us to generate synthetic absorption spectra and then convolve them with the spectrographs' line spread functions to generate final simulated spectra.
The comparisons between the simulated spectra and the observed absorption profiles provide constraints on $U$ and $\FeH$.

We find that most of the lines can be fit with clouds in two photoionized phases, a low-ionization phase (e.g., \ion{Mg}{ii}, \ion{Fe}{ii}, \ion{Mg}{i}) and a high-ionization phase (e.g., \ion{C}{iv}, \ion{N}{iii}, \ion{Si}{iv}).
There is some contribution from both phases in some of the detected lines.
More intricate solutions are also possible, but we focus on the simplest feasible solution.

\section{Low-Metallicity Subsystem} \label{sec:lowzcloud}
At a blueward offset of about $300 \, \kms$ from the associated galaxy, a feature appears in the Lyman series that is not detected in the metal transitions, indicating the presence of low-metallicity gas.
In this section we determine constraints on metallicity for this absorption feature using models based on PIE, CIE, and non-CIE.

\subsection{Photoionization Model}
We first determine a VP fit to the \ion{H}{i} lines to establish $N_\mathrm{HI}$.
Our VP fit can be seen in Table \ref{tab:cloud_parameters} as Clouds \#10 and \#11. We use Cloud \#10 to investigate the upper limit to metallicity of the low-metallicity system, with $\log N_\mathrm{HI} = 18.04 \pm 0.88 \, \mathrm{cm}^{-2}$ and $b_\mathrm{HI} = 15.1 \pm 2.2 \, \kms$.
We determine upper limits for column densities of each metal ionization species by calculating the $EW_\mathrm{R}$ that a barely resolved line would need to have in order to be detected at the $3\sigma$ limit.
A summary of our metal line limits can be found in Table \ref{tab:line}.
We assume $b = 3 \, /kms$ for unresolved lines and $b=10 \, \kms$ for resolved lines.
An unresolved line with $b = 3 \, \kms$ is extremely small, especially considering the much larger Doppler parameters of \ion{H}{i}, but it provides a conservative upper limit for the column densities of the metal lines, which in turn provides a conservative upper limit for $\FeH$.
We use \ion{Mg}{ii} 2796, \ion{C}{iv} 1548, \ion{C}{ii} 1036, \ion{Si}{ii} 1260, \ion{N}{v} 1239, \ion{N}{v} 1242, \ion{O}{vi} 1032, and \ion{O}{vi} 1038.

We cannot use this method for \ion{C}{iii} 977, however, because it is unclear weather some of the absorption found at $v \sim -250 \, \kms$ is associated with this absorption system or whether it is due to a blend.
In order to determine a conservative upper limit to the amount of \ion{C}{iii}, we determine a VP fit for that section of the \ion{C}{iii} 977 line as two clouds (the same number as found for the \ion{H}{i} transitions).
Then, we assume the upper limit of \ion{C}{iii} column density is equal to the largest $N_\mathrm{CIII}$ fit of those two clouds.

\begin{table}
\caption{Low-Metallicity Subsystem Limiting Lines} 
\label{tab:line}
 \begin{tabular}{cccc}
 \hline
	Transition & $W_\mathrm{R}$ [\AA] & $\log (N/\mathrm{cm}^{-2})$ &	b [$\kms$] \\
	\hline
\ion{Mg}{ii} 2796    &0.007    & $<11.24$ & 3.0 \\
\ion{Si}{ii} 1260    &0.021    & $<12.36$ & 3.0 \\
\ion{C}{iv} 1548     &0.014    & $<12.63$ & 3.0 \\
\ion{O}{vi} 1038     &0.01     & $<13.30$ & 3.0 \\
\ion{O}{vi} 1032     &0.01     & $<13.00$ & 3.0 \\
\ion{C}{ii} 1036     &0.01     & $<13.03$ & 3.0 \\
\ion{N}{v} 1239     &0.02     & $<13.16$ & 3.0 \\
\ion{N}{v} 1242     &0.02     & $<13.46$ & 3.0 \\
 \\
\ion{Mg}{ii} 2796    &0.007    & $<11.22$ & 10.0 \\
\ion{Si}{ii} 1260    &0.021    & $<12.22$ & 10.0 \\
\ion{C}{iv} 1548     &0.014    & $<12.56$ & 10.0 \\
\ion{O}{vi} 1038     &0.01     & $<13.23$ & 10.0 \\
\ion{O}{vi} 1032     &0.01     & $<12.93$ & 10.0 \\
\ion{C}{ii} 1036     &0.01     & $<12.96$ & 10.0 \\
\ion{N}{v} 1239     &0.02     & $<13.02$ & 10.0 \\
\ion{N}{v} 1242     &0.02     & $<13.31$ &10.0 \\
\\
\ion{C}{iii} 977      &\dots   & $<13.08$ & 11.9 \\
\hline
\end{tabular}
\end{table}

\begin{figure} %Figure 5
\includegraphics[angle = 00,width=\columnwidth]{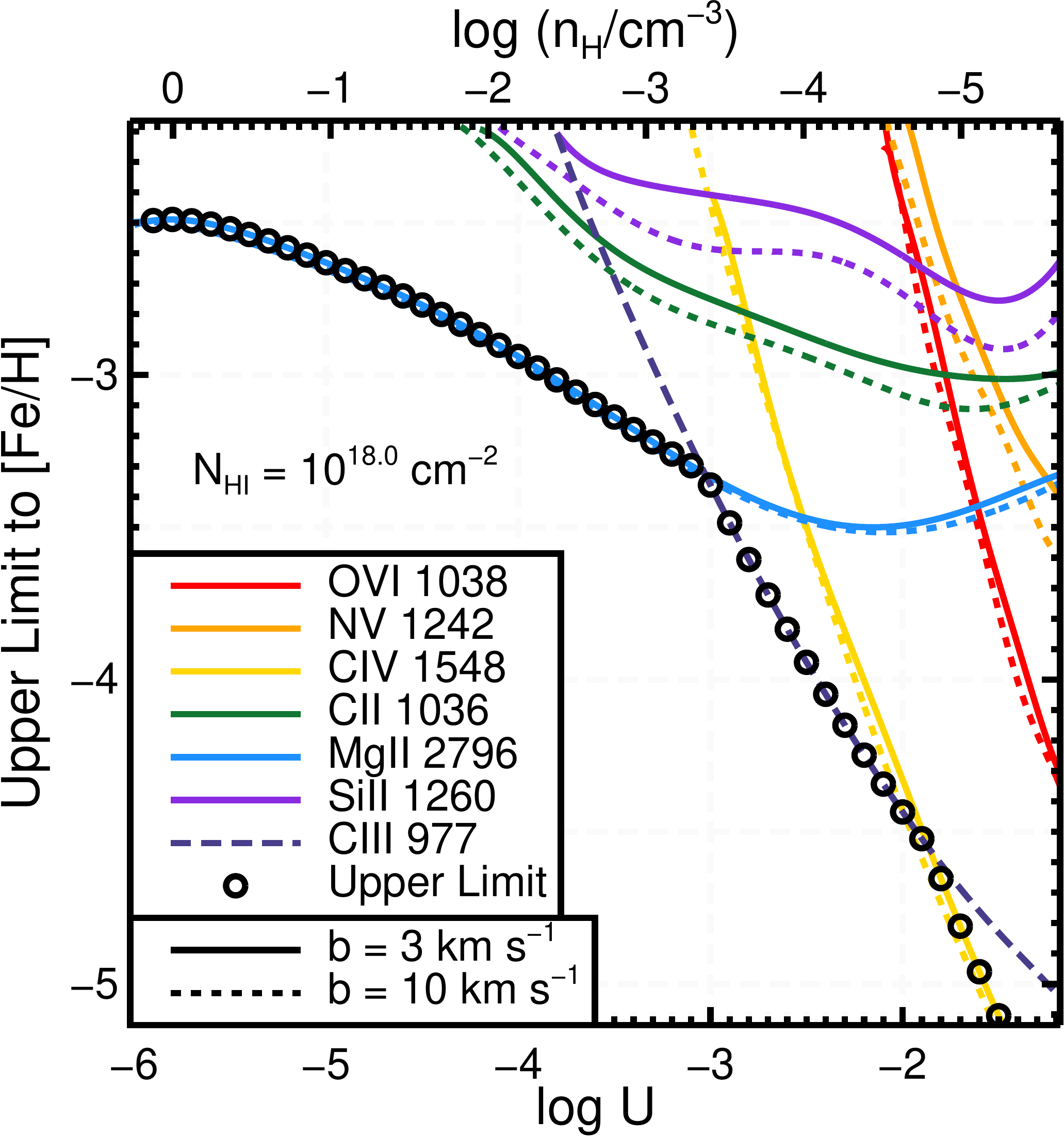}
\caption{Photoionization upper limits of $\FeH$ for a given values of $\log U$ in the low-metallicity subsystem Cloud \#10 based on constraints from various ions.
Each ionization species provides its own upper limit for metallicity based on its maximum $N_\mathrm{X}$ given in Table \ref{tab:line}.
Points above the metallicity of a line will overproduce that line's corresponding ionization species.
By taking the limits from all ionization species concurrently, we determine the most stringent upper limit based on all ions.
\label{fig:PI_logZ_limit}}
\end{figure}

We are unable to constrain $U$ without the presence metal lines.
However, we can determine constraints on $\FeH$ over a given range of values for $U$.
For PIE modeling, the most constraining lines are \ion{Mg}{ii} 2796 when $\log U < -3.0$, \ion{C}{iii} 977 when $-3.0 < \log U < -2.0$, and \ion{C}{iv} 1548 when $-2.0 < \log U$.
We create Cloudy PIE models that match the $N_\mathrm{HI}$ given by our VP fits and also do not exceed the column densities of each metal transition.
Our solutions for $\{U , \FeH\}$ combinations can be found in Figure \ref{fig:PI_logZ_limit}.
For a given value of $U$, Figure \ref{fig:PI_logZ_limit} shows the upper limit of $\FeH$ allowed by the metal lines; if $\FeH$ is above this limit, then the EW of one or more metal lines exceeds the detection threshold.

We find that an extremely conservative metallicity estimate is $\FeH \lesssim -2.5$ (black points in Figure \ref{fig:PI_logZ_limit}).
However, this limit is at a high density of $n_\mathrm{H} \sim 1 \, \mathrm{cm}^{-3}$, so we find more realistic estimate is $\FeH \lesssim -2.9$ (see \S\ref{discussion:lowZ}).

\subsection{Collisional Ionization Model} \label{sec:cie}
\begin{figure} %Figure 6
\includegraphics[angle = 0,width=\columnwidth]{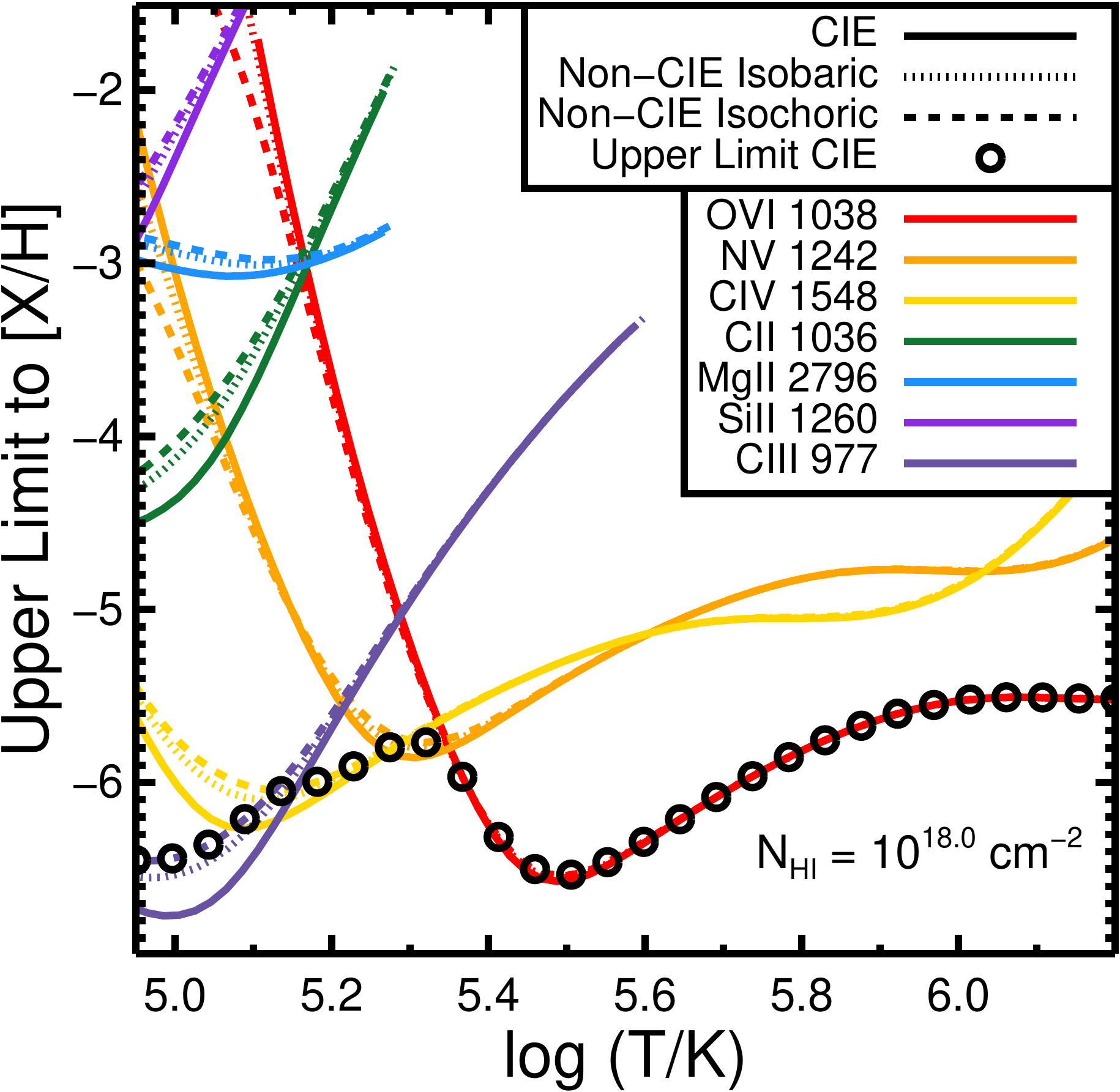}
\caption{Upper limits of metallicity similar to Figure \ref{fig:PI_logZ_limit}, but for collisional ionization models as a function of temperature.
Generally, the CIE model provides lower limits than the isochoric and isobaric models.
We choose to report upper limits of the CIE model, but the metallicity does not change much at all for large $T$.
\label{fig:col_logZ_limit}}
\end{figure}

We adopt models of \citet{gnat+07} (G\&S) that calculate cooling efficiencies and ionization states for low-density, radiatively cooling, dust-free gas devoid of an external radiation field.
In Figure \ref{fig:col_logZ_limit} we present CIE and non-CIE models for the upper metallicity limits provided by each ionization species.
We use the $b = 3 \, \kms$ upper limit column densities established previously (see \S \ref{sec:lowzcloud}).
At high temperatures above $T \gtrsim 10^{5.4}$ K, the non-CIE models give essentially the same upper limits for metallicity as the CIE model.
When temperatures are closer to $T = 10^{4.0}$ K, the collisional ionization models give different metallicity results, but we expect this temperature regime to be PI dominated, so we do not find this temperature regime to be a compelling upper limit for metallicity.

The ionization species most important to constraining metallicity are \ion{O}{vi}, \ion{C}{iii}, and \ion{C}{iv}.
For all temperatures, the limits on metallicity are low, with the maximum being at $\OH = -5.5$ at $T = 10^{6.1}$ K (assuming $T > 10^{5.0}$ K).
Allowing our value of $N_\mathrm{HI}$ to drop to $1 \sigma$ of our VP fit changes the metallicity limit to be $\OH = -4.6$.
Compared to PI models, the collisional models provide a significantly lower metallicity.
However, regardless of which model we adopt, we always find that the low-metallicity subsystem clouds must be extremely metal poor.

\subsection{Alternative EBRs}
We explore the use of an alternative model for the EBR (`KS EBR') by \citet[z=0.48]{khaire+15} that is designed to match the \ion{H}{i} ionization rates for the $z<0.5$ universe \citep{shull+15}.
The upper limit for metallicity of the low-metallicity subsystem clouds provided by the KS EBR does not differ appreciably from the HM12 EBR limit.
When $\log U \gtrsim -2.3$, the KS EBR upper limit tends to increase to $\FeH \lesssim -5.1$, but the absolute upper limit of metallicity for any ionization parameter is still $\FeH < -2.5$ as with the HM12 EBR (see Figure \ref{fig:PI_logZ_limit}).

\subsection{Alternative H I Column Density} \label{sec:alternative_hi}
Since $N_\mathrm{HI}$ is extremely important for determining the upper limit to the low-metallicity subsystem, we explore the validity of alternative VP solutions for \ion{H}{i}.
Such solutions may be possible because of uncertainties in the wavelength calibration of the COS data, and thus are considered here.
We are unconcerned with VP solutions in which $N_\mathrm{HI}$ is larger than the values we present for Clouds \#10 and \#11 in Table \ref{tab:cloud_parameters}.
This is because for column densities larger than those we present in the table, the upper limit for metallicity is \textit{less} than the derived upper limit in Figure \ref{fig:PI_logZ_limit}.
This means that larger column density solutions are even more conservative.

For column densities less than those we present in the table, the upper limit for metallicity scales roughly with $N_\mathrm{HI}$.
If $N_\mathrm{HI} = 17.0 \, \cmcm$ for the larger cloud in the low-metallicity subsystem, then the upper limit for metallicity would scale by about $1.0$ dex to $\mathrm{[Fe/H] < -1.5}$.
Although this value of column density is only $1.2\sigma$ away from the best fit, it should be noted that it does not visually provide a satisfactory fit.
We suspect that the error provided by the VPFIT software for $N_\mathrm{HI}$ may be too large (see \S \ref{sec:caveats}).
For a point of reference, we include a $N_\mathrm{HI} = 10^{17.0} \, \cmcm$ variation of Cloud \#10 in Figure \ref{fig:cld11}.

Nonetheless, we explore a different VP fit solution in which we force $N_\mathrm{HI} = 17.0$ for Cloud \#10 to determine if it can be realistically consistent with the data.
Looking at only the Lyman lines, we were able to determine a separate 2-cloud VP fit to the low-metallicity subsystem in which $N_\mathrm{HI} = 17.0 \, \cmcm$ that is different from the fit we have shown.
However, this solution requires large velocity shifts in the detected transitions in the opposite direction as the velocity shifts provided in Figure \ref{fig:transplotq_lowu}.
Essentially, this VP fit requires the data in the Lyman absorption to be shifted to the red.
In such a case, the velocity offsets require the low-ionization phase to have strongly different solutions.
It is possible to obtain a moderately good fit to the data with the low-ionization lines if a strong metallicity gradient is imposed across the clouds in the low-ionization phase.
This is because shifting the data redward means that the low-redshift components of the low-ionization phase model must have smaller \ion{H}{i} column densities compared to the fits provided in Table \ref{tab:cloud_parameters} while the high-redshift components must have larger \ion{H}{i} column densities.
The net result is the low-redshift clouds would have much stronger metallicity solutions ($\mathrm{[Fe/H] \sim +0.6}$) while the high-redshift clouds would have much weaker metallicity solutions ($\mathrm{[Fe/H] \sim -2.2}$).
This fit requires the metallicity solutions of the six low-ionization phase clouds to span 2.8 orders of magnitude, which seems excessive for gas originating within the same phase.
We do not suspect that a model with $N_\mathrm{HI} = 10^{17.0} \, \cmcm$ is a viable solution because it seems contrived that there would be a strong metallicity gradient across the low-ionization clouds.
Additionally, the derived metallicity within this model for some of the low-ionization material is still extremely low ($\FeH \lesssim -2.2$), meaning it is should not be any more likely than low-metallicity material in the blueward absorption subsystem.

\subsection{Deuterium} \label{sec:deuterium}
Deuterium is detected in the low-metallicity subsystem.
This is most pronounced directly blueward of the \ion{H}{i} $\mathrm{Ly\,\alpha}$ transition.
Unfortunately, determining a precise column density of D is difficult due to the SNR of the data.
Within our fit shown in Figures \ref{fig:transplotq_lowu} and \ref{fig:transplotq_highu}, we present an abundance ratio of $\mathrm{D/H = -4.6}$ that is expected from big-bang nucleosynthesis (BBN) \citep{cyburt+16}.
Our model agrees with what we expect from BBN, which corroborates the idea that the low-metallicity subsystem is pristine gas with low-metallicity.
However, our data do not have a SNR with sufficient quality to measure the D/H abundance precisely, so we are unable to differentiate whether the detected deuterium is necessarily from BBN or whether it is from other types of material similar to the ISM in the Galactic disc, which has been shown to have variations between $ -5.3< \mathrm{D/H} < -4.65$ \citep{linsky+06}.

\section{Low-Ionization Phase} \label{sec:lowucloud}
The low-ionization phase is the primary source for absorption in \ion{Mg}{i}, \ion{Mg}{ii}, \ion{Fe}{ii}, \ion{Ca}{ii}, \ion{Si}{ii}, \ion{C}{ii}, \ion{C}{iii}, and the Lyman series.
Low-ionization clouds may also partially contribute to absorption in \ion{C}{iv}, \ion{Si}{iv}, and \ion{N}{iii}.
Due to the high SNR of the UVES data, we use the \ion{Fe}{ii} transitions as the starting point for our modeling.
We find six clouds that are generally characterized by $\FeH \sim \{-1.0,+1.0\}$ and $\log U \sim -3.5$ (see Table \ref{tab:cloud_parameters}).
This phase is dominated by photoionizaton with temperatures too low for collisional ionization to influence the line profiles.

The lower limit for metallicity for each cloud can be determined through more than one method, depending on the specific cloud.
When considering the low-ionization phase as a whole, the two most blueward clouds (\#1 and \#2) and the most redward cloud (\#6) are important since those clouds most affect the wings of the Lyman series lines and cannot produce absorption exceeding that observed.
The central clouds (\#3, \#4, \#5) are in a regime where the \ion{H}{i} lines are saturated and the components are blended with one another, thus \ion{H}{i} does not have a meaningful limit on metallicity here and other transitions are necessary to constrain metallicity.

\begin{figure*} %Figure 7
\includegraphics[angle = 0,width=\textwidth]{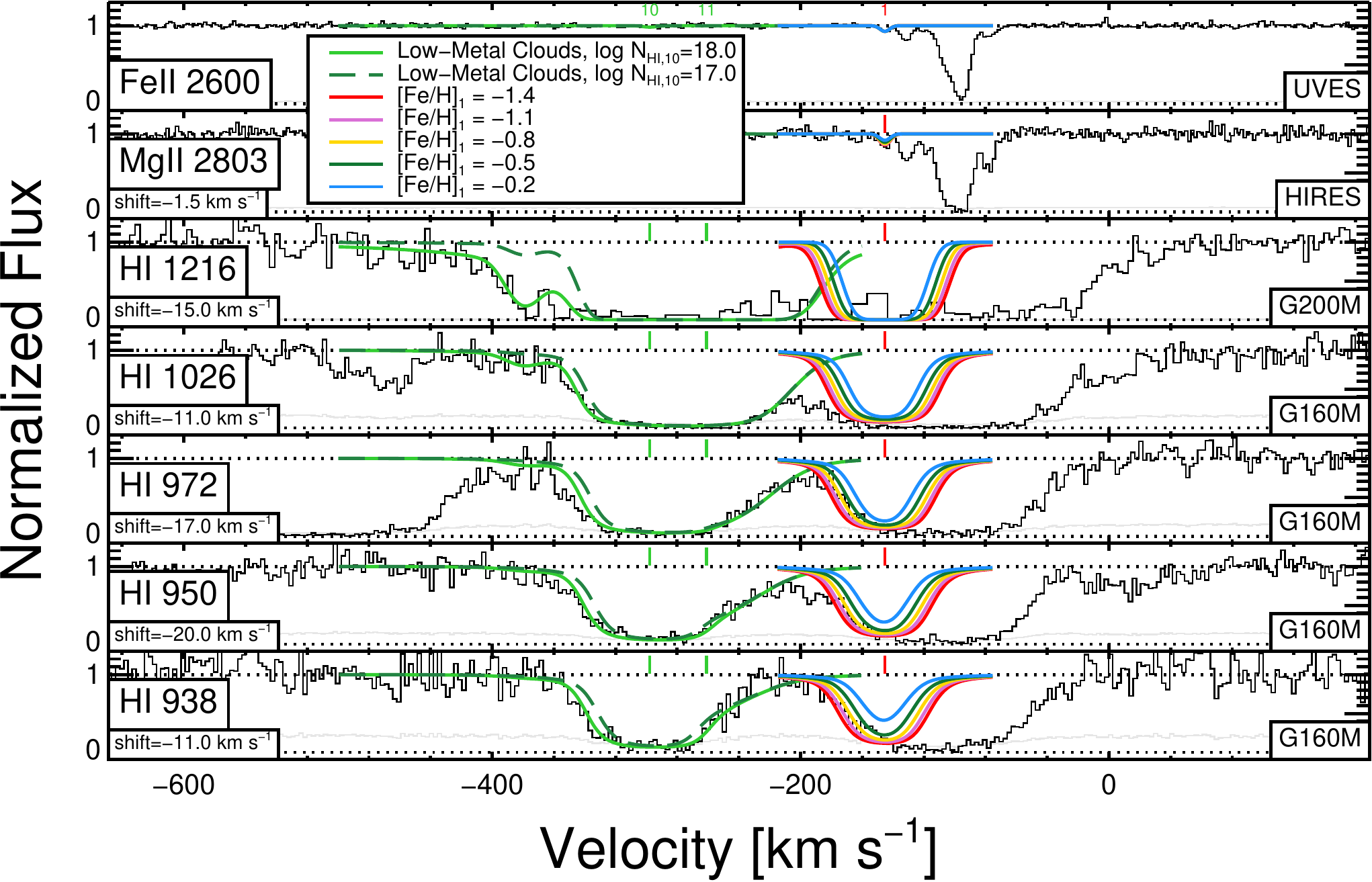}
\caption{Demonstration of how changing $\FeH_1$ affects the fit for Cloud \#1 in the model.
Note that several values of $\FeH_1$ are acceptable, but some clearly overproduce the Lyman series.
Underproducing the Lyman series here is somewhat less of a problem, since the blueward component of the low-ionization phase can be fit with a superposition of Clouds \#1 and \#2 simultaneously.
We find that Cloud \#1 has a metallicity limit of $-1.2 \lesssim \FeH_1$.
We are certain of the velocity shifts because the blueward low-metallicity subsystem clouds (shown in green) must be fit simultaneously.
\label{fig:cld11}}
\end{figure*}

We find that Cloud \#1 has a metallicity limit of $-1.2 \lesssim \FeH_1$ and an ionization parameter limit of $\log U_1 \lesssim -3.2$.
The metallicity limit is determined by the Lyman series lines (see Figure \ref{fig:cld11} for a demonstration), while the ionization limit for is provided by the \ion{Si}{ii} and \ion{Mg}{ii} transitions; when $-3.2 \leq \log U_1$, then \ion{Si}{ii} and \ion{Mg}{ii} are overproduced by the model.
In principle, an even lower limit on $U_1$ could be derived from \ion{Si}{ii}, but this is uncertain because \ion{Si}{ii} is not sensitive to changes in $U_1$ in this regime.
We are unable to place an upper limit on $\FeH_1$ because the Lyman series lines could be explained using only Cloud \#2 or by a contribution from the higher ionization phase.
To match the left wing of \ion{H}{i} with the contribution from Cloud \#1, a value of $\FeH_1 \sim -0.8$ is consistent, which we use in our ``best'' fit.

Cloud \#2 has limits of $-1.8 \lesssim \FeH_2 \lesssim -0.8$, and $-3.9 \lesssim \log U_2 \lesssim -3.7$.
The lower limit on $\FeH_2$ comes from the requirement that the blueward wing of the Lyman series lines is not overproduced by this cloud.
From \ion{H}{i} alone, we cannot find an upper limit for $\FeH_2$, because a superposition of Clouds \#1 and \#2 likely causes the blueward wing of the line.
The ratio of \ion{Fe}{ii} to both \ion{Mg}{ii} and \ion{Si}{ii} is sensitive to changing $U$ and $\FeH$.
The depths of the two components of the \ion{Mg}{ii} doublet appear to be slightly inconsistent with what is expected from atomic physics within the data, perhaps due to small continuum fitting uncertainties; we emphasize the fit to the \ion{Mg}{ii} 2803 line in our modeling.
The lower limit of $-3.9 \lesssim \log U_2$ ensures that \ion{Mg}{ii} 2803 is not overproduced, while the upper limit of $\log U_2 < -3.7$ avoids overproduction of \ion{Si}{ii}.
The upper limit on metallicity is less certain but is determined by \ion{Si}{ii}, for which values greater than $-0.8 < \FeH_2$ overproduce \ion{Si}{ii} relative to \ion{Mg}{ii} in the model; for larger values of metallicity, either \ion{Mg}{ii} is underproduced \ion{Si}{ii} is overproduced.

Cloud \#3 has limits of $-3.6 \lesssim \log U_3 \lesssim -2.3$ and an unrestricted metallicity limit of $-3.0 \lesssim \FeH_3$.
Unfortunately, all covered \ion{H}{i} lines associated with this cloud are blended with the contributions from other clouds, so \ion{H}{i} cannot assist in the modeling.
The lower limit of $U_3$ is primarily constrained by \ion{Mg}{ii} (relative to \ion{Fe}{ii}), as lower values of $U_3$ tend to underproduce \ion{Mg}{ii} in the model.
In the very low metallicity regime, the ionization parameter can be as large as $\log U_3 \lesssim -2.3$, however the densities are unrealistically low so that cloud sizes become prohibitively large at greater than 1 Mpc.
For the permitted values of $\log U_3$, the observed \ion{C}{iv} is underproduced in the model, thus requiring a separate high-ionization phase (see \S \ref{sec:highucloud}).

Our model constraints for Cloud \#4 and Cloud \#5 are affected by apparent inconsistencies between the \ion{Mg}{ii} 2796 and 2803 lines at $v = -110 \, \kms$ and $-90 \, \kms$ similar to Cloud \#2.
This problem may simply reflect noise in the data, may be due to small ~problems with data reduction, or may be a blend with an unidentified line contributing at that position of the 2796 transition.
In what follows we favor the \ion{Mg}{ii} 2803 transition in constraining our models.

Cloud $\#4$ is found to have a high metallicity of $-0.1 \lesssim \FeH_4$ and an ionization parameter in the narrow range $-3.1 \lesssim \log U_4 \lesssim -2.9$.
For this cloud, in order to avoid overproduction of \ion{Mg}{i} (relative to \ion{Fe}{ii}), the metallicity must either be solar or higher, or it must be considerably lower, $-2.0 \lesssim \FeH_4$ (i.e.models with metallicities between $-2.0 \lesssim \FeH_4 \lesssim -0.1$ overproduce \ion{Mg}{i}).
However, models within the low metallicity regime generally underproduce the \ion{Mg}{ii}, so the high metallicity regime is favored.
(An additional extremely narrow region of parameter space at $\FeH_4 \sim -1.9$ and $\log U_4 \sim -2.3$ is possible, but the cloud size is close to a prohibitive size of 1~Mpc, so we reject this special case.)
At $-0.1 \lesssim \FeH_4$ the ratio of \ion{Mg}{ii} to \ion{Fe}{ii} constrains the ionization parameter to be in a narrow range close to $\log U_4 = -3.0$.
For the favored model, the \ion{C}{iv} is underproduced, so as with Cloud \#3, a separate high-ionization phase is required.

Cloud \#5 is poorly constrained because it is not close to the edge of the Lyman series line profiles and because it is optically thick over much of  parameter space.
In particular, the relationships between \ion{Fe}{ii}, \ion{Mg}{ii}, \ion{Si}{ii}, and \ion{Mg}{i} are complex.
Solutions that match the observed \ion{Mg}{i} exist in two ranges of metallicity, at roughly $\FeH_5 \sim -0.25$ and $\FeH_5 \sim -1.3$.
For the higher metallicity range,  values of $\log U_5 \lesssim -3.2$ are needed in order for \ion{Mg}{ii} and \ion{Si}{ii} to not be overproduced in the model, while within the lower metallicity range values of $\log U_5 \lesssim -2.5$ are permitted.
Our `best' model greatly overproduces \ion{Ca}{ii} ($N_\mathrm{CaII} = 12.15$).
A single-cloud VP fit to the \ion{Ca}{ii} 3935 data gives $N_\mathrm{CaII} = 11.41 \pm 0.04$.
This discrepancy between our model and the data is a difference of 0.74 dex, which indicates a deviation from the solar abundance pattern (see \S \ref{sec:caveats}).

Cloud \#6 is constrained to have $-1.4 \lesssim \FeH_6$ and $-3.7 \lesssim \log U_6 \lesssim -3.2$.
Below the metallicity limit, the Lyman series lines are overproduced in the model.
When $U_6$ is below the allowed range, \ion{Mg}{ii} is underproduced in the model for all values of $\FeH_6$ relative to \ion{Fe}{ii}; when $U_6$ is above the allowed range, the \ion{Mg}{ii} and \ion{Si}{ii} transitions are overproduced.
For solutions in the allowed range of $U_6$, \ion{Mg}{i} is consistent, but \ion{C}{iv} requires a separate phase as with Clouds \#3 and \#4.

In summary, several of the low-ionization clouds have higher metallicities than the low-metallicity subsystem clouds.
The density of the clouds are similar, and our `best' fit range for the six clouds is at $n_\mathrm{H} \sim \{10^{-2.7},10^{-1.8}\} \, \mathrm{cm}^{-3}$.
The clouds are narrow with Doppler parameters of $b_\mathrm{Fe} \lesssim 5 \, \kms$.
Table \ref{tab:cloud_parameters} gives these values in greater detail as well as the parameters of our fits for each cloud, and the `best' model which we use to create Figure \ref{fig:transplotq_lowu} and Figure \ref{fig:transplotq_highu}.

\section{High-Ionization Phase} \label{sec:highucloud}
Based on constraints from the low- and intermediate-ionization transitions, the observed \ion{C}{iv} and \ion{O}{vi} transitions cannot both be fully produced by the low-ionization clouds.
Of the three high-ionization species (\ion{C}{iv}, \ion{N}{v}, and \ion{O}{vi}), \ion{C}{iv} provides the best starting point for modeling.
\ion{N}{v} is weakly detected, and the \ion{O}{vi} doublet falls along a part of the spectrum affected by blends, so it is not possible to measure the \ion{O}{vi} column density well.
The \ion{C}{iv} doublet is best fit with a 2-cloud model with one broad redward cloud (Cloud \#8, with $N_\mathrm{CIV} \sim 10^{14.1} \, \cmcm$ and $b \sim 24 \,\kms$) and a narrower blueward cloud (Cloud \#7, with $N_\mathrm{CIV} \sim 10^{13.4} \, \cmcm$ and $b \sim 7 \, \kms$), as seen in Figure \ref{fig:transplotq_highu}.
It is possible that Cloud \#8 is composed of two smaller clouds with similar Doppler parameters ($b \sim 16$), but this only seems to be apparent in \ion{C}{iv} 1548 alone, and does not appear in \ion{C}{iv} 1551.
For our `best' fit, we opt for the 2-cloud model.

\ion{C}{iv} by itself can provide a lower limit for the metallicity of the high-ionization phase (see \S \ref{sec:method}).
For a given combination of $U$ and $\FeH$, as $N_\mathrm{HI}$ increases $N_\mathrm{CIV}$ increases as well until $N_\mathrm{HI} \sim 10^{17.5} \, \mathrm{cm}^{-2}$, at which point the material becomes optically thick and self-shielding starts to limit the amount of \ion{C}{iv}.
At this point, $N_\mathrm{CIV}$ plateaus at a constant value and will not increase any further, even for large values of $N_\mathrm{HI}$.
Thus, for a given combination of $U$ and $\FeH$, there is a maximum value of $N_\mathrm{CIV}$ that can arise.
To generalize, as smaller values of $\FeH$ are adopted, the maximum allowed $N_\mathrm{CIV}$ decreases.
This means that for the observed value of $N_\mathrm{CIV}$ there exists a lower limit on $\FeH$ below which it is impossible to produce enough \ion{C}{iv} for any value of $U$.
For Cloud \#7, with $N_\mathrm{CIV} \sim 10^{13.3} \, \cmcm$, a value of $\log U_\mathrm{7}=-2.2$ produces a constraint of $-3.3 < \FeH_\mathrm{7}$.
For Cloud \#8, with $N_\mathrm{CIV} \sim 10^{14.0} \, \cmcm$, a value of $\log U_\mathrm{8}=-2.2$ produces a constraint of $-2.6 < \FeH_\mathrm{8}$.
For higher ionization parameters in both clouds, the metallicities could be lower, but then the cloud sizes would exceed a prohibitive size of 1 Mpc.
The lower limits for $U$ (-2.9 < $\log U_\mathrm{7}$ and $-2.9 < \log U_\mathrm{8}$) are extremely conservative, constrained only by the requirement that the high-ionization phase not overproduce the \ion{Mg}{ii} transitions, which were already well-modeled by the low-ionization phase (see \S \ref{sec:lowucloud}).
Additionally, we find upper limits for $U$ at $\log U_\mathrm{7} < -1.4$ and $\log U_\mathrm{8} < -1.5$ due to \ion{O}{vi}; above these values, the \ion{O}{vi} transitions are overproduced.

Beyond these loose constraints on metallicity, there are limited other constraints on the properties of the high-ionization phase.
For Cloud \#7, the lack of detected \ion{N}{v} ($N_\mathrm{NV} < 10^{12.7} \, \cmcm$) places a constraint of $\log U_\mathrm{7} \lesssim -1.7$ for a solar abundance pattern.
Similarly, the low column density of detected \ion{N}{v} in Cloud \#8 ($N_\mathrm{NV} < 10^{13.4}$) places a constraint of $\log U_\mathrm{8} \lesssim -1.7$.
These values of $U_7$ and $U_8$ do not come close to overproducing \ion{O}{vi} in a region of the spectrum heavily contaminated with blends.
If instead, we assumed that the maximum amount of \ion{O}{vi} is associated with the \ion{C}{iv} clouds then $\log U_7 \lesssim -1.4$ and $\log U_8 \lesssim -1.5$, which would overproduce the \ion{N}{v} for a solar abundance pattern.
This is not unusual for the high-ionization phase of other QAL systems in the literature (see discussion in \S \ref{discussion:nitrogen}).

It is also possible that either the \ion{C}{iv} or the \ion{O}{vi} could arise from collisionally ionizied gas.
However, if this be the case, solutions that produce both \ion{C}{iv} and \ion{O}{vi} at $T \sim 10^{5.3}$ K ($b_\mathrm{C} \gtrsim 16 \, \kms$) also overproduce \ion{N}{v} similar to the PIE model.

In summary, a high-ionization phase is needed to explain the observed \ion{C}{iv} absorption.
It could be photoionized or collisionally ionized, but it is not possible to place strong constraints on either the density or temperature because of confusion in the region of the spectrum covering \ion{O}{vi}.
If photoionized, the metallicity of the high-ionization gas must be $\FeH > -2.5$, which exceeds that of the low-metallicity subsystem clouds, but otherwise $\FeH$ is unconstrained.

\section{Intermediate-Ionization Phase} \label{sec:intermediateucloud}
Some absorption on the redward side of the \ion{H}{i} 1216 transition and the \ion{C}{iii} 977 transition cannot be explained via the low-ionization or high-ionization phase.
We fit this absorption with Cloud \#9, which has an intermediate-ionization phase.
\ion{C}{iii} absorption has occurred in previous systems where little or no other absorption is detected besides \ion{H}{i} \citep[e.g.,][]{lacki+10}.
Within a narrow range of values for $U$, our Cloudy models produce sufficient \ion{C}{iii}, but produce little \ion{C}{iv}.
This cloud strongly produces absorption in only two transitions, but we are still able to constrain its physical parameters well.
We fit Cloud \#9 with $-0.1 \lesssim \FeH_9$ and $-2.8 \lesssim \log U_9 \lesssim -2.1$.
The metallicity lower limit is provided by the \ion{H}{i} transitions, since this cloud must not produce much \ion{H}{i} absorption beyond the \ion{H}{i} 1216 transition.
The lower limit of ionization parameter is also provided by the \ion{H}{i} transitions, since below $\log U_9 = -2.8$, hydrogen tends to be overproduced regardless of metallicity.
The upper limit of ionization parameter is provided by the \ion{C}{iv} transitions, since it is necessary for this cloud to produce little or no \ion{C}{iv} absorption.

\section{Modeling Caveats} \label{sec:caveats}
Our results for the parameters of our clouds are based heavily on our VP fits of the \ion{Fe}{ii}, \ion{C}{iv}, and \ion{H}{i} lines.
Therefore, any errors we may have made in our VP fits will propagate into our models.
In this section we investigate the robustness of our findings to these problems.

Our Cloudy modeling focuses on $N_\mathrm{X}$, $\FeH$, and $U$, but is not dependent on the Doppler parameter $b_\mathrm{X}$.
However, if values for $b_\mathrm{X}$ in our VP fits are inaccurate, this will alter the fit for $N_\mathrm{X}$.
We investigate solutions in which we artificially force our VP fit to have a smaller $b_\mathrm{X}$.
Lowering the velocity spread of a metal in our VP fit tends to require a larger metal column density to maintain the same $EW_\mathrm{R}$ for the line, which tends to decrease the metallicity solution in Cloudy models for the low-ionization and high-ionization phases.

We perform a case study for this effect on Cloud \#4 from the low-ionization phase.
We artificially decrease the Doppler parameter to $b_\mathrm{Fe} = 1.00 \, \kms$ instead of the best fit of $b_\mathrm{Fe} = 2.79 \, \kms$.
The curve of growth for \ion{Fe}{ii} is particularly susceptible to changes in $b$ at these low velocities, so in order to keep the same $EW_\mathrm{R}$ we are forced to increase the column density from $N_\mathrm{FeII} = 10^{12.60} \, \cmcm$ to $N_\mathrm{FeII} = 10^{13.90} \, \cmcm$.
We scaled the other ionization species' Doppler parameters similarly to provide different constraints on column density for each ionization species.
With the increased column densities, we achieve a lower limit of $-1.3 \lesssim \FeH_4$, which is significantly lower than our initial lower limit of $-0.1 \lesssim \FeH_4$.
However, the $-1.3 \lesssim \FeH_4$ limit is an extreme solution around $\log U \sim -1.6$, where most of the \ion{C}{iv} and \ion{O}{vi} is produced in the same phase as the \ion{Mg}{ii} and \ion{Mg}{i}, and $N_\mathrm{HI}$ is dangerously close to overproducing the Lyman series.
Decreasing metallicity further will underproduce the \ion{Mg}{ii} lines and overproduce the Lyman series.
We conclude that this solution is untenable, but even if it is to be adopted, it still has a metallicity which is clearly higher than that of the low-metallicity subsystem clouds.

The \ion{Ca}{ii} feature is overpredicted in our low-ionization phase model, but this is expected based on previous systems.
A VP fit to the data indicate $N_\mathrm{CaII} = 10^{11.4} \, \cmcm$ near Cloud \#5.
Using our best fit for Cloud \#5 of $N_\mathrm{HI} = 10^{17.2} \, \cmcm$, this gives us a ratio of $\log \frac{\mathrm{CaII}}{\mathrm{HI}} = -5.8$.
This is close to the expected value $\log \frac{\mathrm{CaII}}{\mathrm{HI}} = -5.5$ predicted from a (biased) fit of seven similar systems QAL systems \citep{richter+11}, and similar to Milky Way disc and halo observations \citep{wakker+00}.
This indicates dust depletion in our system similar to these environments.

As a note of caution, our upper limit on metallicity for the low-metallicity subsystem is highly dependent on our fit for $N_\mathrm{HI}$.
Our VP fit for Cloud \#10 gives a value of $N_\mathrm{HI} = 10^{18.0 \pm 0.9} \cmcm$, which has a significant error.
A smaller value of $N_\mathrm{HI}$ will tend to have a higher upper limit for metallicity, so we investigate the change in our results when we decrease $N_\mathrm{HI}$ by roughly $1 \sigma$.
When we decrease $N_\mathrm{HI}$, it is clear upon visual inspection that the blueward edge of the Lyman series is noticeably underproduced (see Figure \ref{fig:cld11}).
Therefore, we suspect that the error provided by the VPFIT software may be inaccurate in this instance.
The VPFIT authors state that `error estimates provided by VPFIT range from being reasonable to fairly high overestimates' \citep[VPFIT manual v11.1,][]{vpfit14}.
Nonetheless, if we adjust the \ion{H}{i} column density to be $1 \sigma$ higher based on the given error, the upper limit for metallicity changes to $\FeH_{10} < -1.8$.
Although this new upper limit to metallicity is higher than our original model, it is still quite low, and does not greatly alter our interpretation that the metallicity for the low-metallicity subsystem is remarkably low, nor does it alter our interpretation of the possible origin of the subsystem.

We investigate the effect of using Cloudy's `HM05' EBR \citep{ferland+98} in our model.
Although the HM05 EBR is a softer spectrum than the HM12 EBR, at around the 1–3 Ryd level, the HM05 EBR contains more photons than the HM12 EBR. The result is that the solutions for the low-ionization clouds tend to have lower values for $\log U$ while the high-ionization clouds tend to have higher values.
We find that the low-ionization clouds tend to have a value of $\log U$ that is  $\sim 0.4$ lower with the HM05 EBR compared to the the HM12 EBR, while the high-ionization clouds tend to have a value of $\log U$ that is $\sim 0.3$ higher with the HM05 EBR.
The metallicity is only slightly affected by EBR, however, with the low-ionization clouds having $\sim 0.1$ lower values for $\FeH$ and the high-ionization clouds having $\sim 0.1$ higher.
This metallicity difference agrees with the findings of \citet[][, appendix]{wotta+16}, who find an average decrease of $\sim 0.3$ dex when using the HM05 EBR, and \citet{zahedy+2019} who find an average decrease of $\sim 0.3$ dex for low-ionization gas with $\log N_\mathrm{HI} \sim 16 \cmcm$.
Regarding the low-metallicity subsystem, since the photoionization upper limit for metallicity is clearly within the low-ionization regime (Figure \ref{fig:PI_logZ_limit}), then using the HM05 EBR would tend to \textit{lower} the upper limit for metallicity.
A more thorough investigation of the changes to metallicity a high-ionization would require looking at how subtle changes within all the ionization species; for example, changes in \ion{Mg}{ii} and \ion{Mg}{i} may slightly alter the metallicity limits for the low-ionization clouds.
In short, although HM05 alters the exact values for $\FeH$, there is still a clear trend in which the low- and high-ionization phases have much higher metallicity than the low-metallicity subsystem, so the key points of our discussion (see \S \ref{sec:discussion}) should not be affected.

\section{Summary of Results} \label{sec:summary}
\begin{figure*} %Figure 8
\centering
\includegraphics[angle = 270,width=\textwidth]{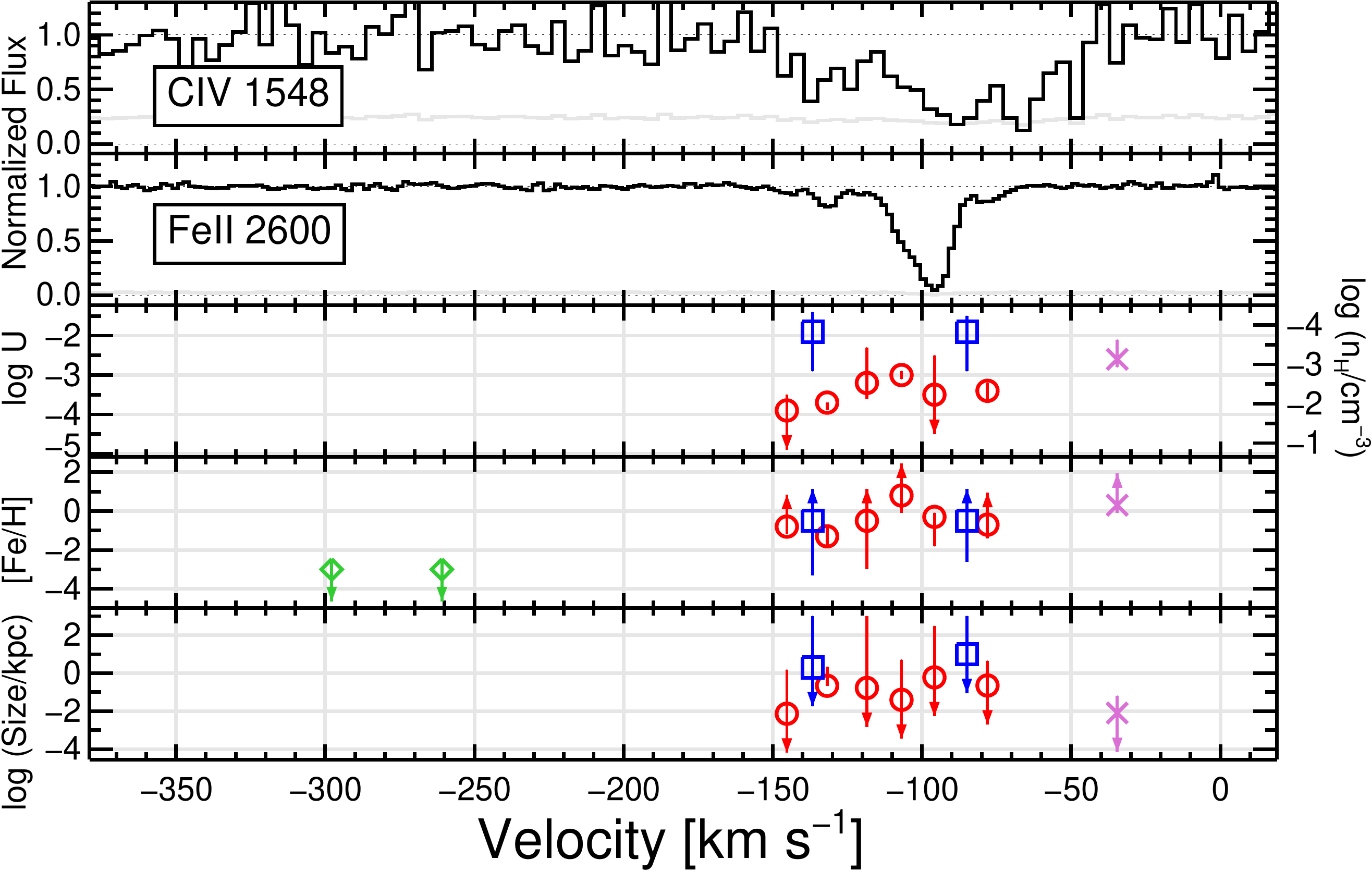}
\caption{Summary of photoionization model results.
The low-metallicity subsystem clouds are labelled with green diamonds, low-ionization clouds with red circles, high-ionization clouds with blue squares, and the intermediate-ionization cloud with a magenta cross.
The geometrical symbol is centred on our `best' fit while the `error bars' correspond to the upper and lower limits we place on the specific cloud.
Some of these limits are narrow, and so the error bar may appear smaller than the geometrical symbol (e.g. $\log U$ for low-ionization Cloud \#2 and \#4).
Downward pointing arrows correspond to a lack of a lower limit while upward pointing arrows correspond to a lack of an upper limit.
The low-metallicity subsystem clouds are not shown in the $\log U$ or size plots because they have no upper or lower limits.
Some upper limits for size are placed at 1 Mpc, but this limit is only based on our assumption that clouds larger than this are unlikely rather than a physical limitation in the Cloudy modeling.
See Table \ref{tab:cloud_parameters} for a detailed list of values used in this figure.
\label{fig:uz_0454}}
\end{figure*}

We consider photoionization and collisional ionization models of the $z=0.4833$ absorption system towards Q0454-220, constrained by spectra from \hstcos, \hststis, \vltuves, and \keckhires.
These spectra cover a range of ionization states, with the most useful constraints provided by \ion{Mg}{ii}, \ion{Fe}{ii}, \ion{C}{iv}, \ion{O}{vi}, and the Lyman series.
The absorption system, shown in Figure \ref{fig:transplotq_lowu} and Figure \ref{fig:transplotq_highu}, spans a $\sim 300 \, \kms$ velocity range, with two broad, separated absorption regions in the Lyman series.
The blueward absorption region has no detected metal absorption lines, despite the fact that the absorption indicates $N_\mathrm{HI} \sim 10^{18} \cmcm$, suggesting remarkably low metallicity.
In contrast, the redward absorption region has only slightly stronger Lyman series absorption, yet it has strong absorption detected in many metal transitions.
See Figure \ref{fig:uz_0454} for a summary of constraints on metallicity, ionization parameter, and density.

The system has an associated galaxy at $z = 0.48382$ with a luminosity of $L_B = 0.9 L_B^{*}$ at an impact parameter of $D = 108.0 \pm 0.6 \, \mathrm{kpc}$.
The kinematics of the metal-line components are consistent with a blueward extension of the galaxy rotation curve \citep{kacprzak+10a,kacprzak+15}, however the low-metallicity subsystem exists well beyond that velocity.

We begin by summarizing model constraints on the redward part of the profile, for which metals are detected.
This region can be fit with six components, based on the \ion{Fe}{ii} lines (the low-ionization phase of the enriched gas).
The exact constraints on the six components differ, but they generally have $\FeH \sim -0.5$ with densities of the order of $n_\mathrm{HI} \sim  10^{-2.3} \, \mathrm{cm^{-3}}$.
The outer two clouds have the best constraints on metallicity, with lower limits of $-1.2 < \FeH$ for Cloud \#1 and $-1.4 < \FeH$ for Cloud \#6.

The high-ionization transitions detected in this same velocity range can be modeled with either a one or two component fit to the \ion{C}{iv} doublet (the high-ionization phase of the enriched gas).
Because the regions of the spectrum covering {\ion{N}{v}} and {\ion{O}{vi}} are confused, the constraints on this phase are not strong.
It could be photoionized (with density lower than the low-ionization phase) or collisionally ionized.
For a photoionized cloud, the lower limit to the metallicity of the dominant high-ionization cloud is $-2.5 < \FeH$, and the upper limit is unbounded.

The blueward, low-metallicity subsystem has a velocity spread of $\sim 150 \, \kms$ and can be fit with two clouds with $N_\mathrm{HI} = 10^{18.0} \, \cmcm$ and $10^{15.3} \, \cmcm$.
In the absence of metal-line detections, its density is not constrained.
For each density, we derive an upper limit on metallicity.
If it is high density, then it must have $\FeH < -2.5$ in order to produce undetected \ion{Mg}{ii}, and if it is lower density then $\FeH < -3.0$ in order not to produce \ion{C}{iii} absorption.
Regardless, this is substantially (one and a half orders of magnitude) below the metallicities of the redward metal-enriched \ion{H}{i} feature.
With G\&S CIE and non-CIE models, upper limits on the metallicity are even lower.

Considering these constraints from photoionization and collisional ionization models, as well as the properties of the absorbing galaxy at an impact parameter of  $D = 108.0 \pm 0.6 \mathrm{kpc}$,  we now compare this absorber to other absorbers, and discuss its possible origins.

\section{Discussion} \label{sec:discussion}
\subsection{Physical Geometry and Origin}  \label{sec:discussiongal}
In order to constrain the possible sources of the absorption, we must first constrain the three dimensional orientation of the galaxy with respect to the observer.
Without this knowledge, then there is a degeneracy with the sign of disk inclination angle, which indicates how the galaxy disk tilts with respect to the plane of the sky.
From the observer's perspective, switching the sign flips the near side and the far side of the galaxy disk, which could change how we interpret our results.
Using the methods of \citep{Ho+20}, we use the wrapping direction of spiral arms and the direction of disk rotation from the rotation curve to determine which way the galaxy disk is orientated on the sky.
In Figure~\ref{fig:diagram}, we can see that the spiral arms are wrapped in a counterclockwise direction.
The direction of rotation using the galaxy rotation curve from \citet{kacprzak+10b} is also shown in Figure~\ref{fig:diagram}. 
This combination of counterclockwise spiral arm wrapping, and the direction of rotation indicates that the galaxy disk is away from the observer along the quasar line-of-sight (see \citet{Ho+20} Figure~2). 

We show a simple diagram of the system in Figure~\ref{fig:diagram}.
The diagram is constructed so that the reader is looking `from the side,' with the observer on the left and the QSO on the right.
The galaxy appears slightly face-on to the observer with $i = 42.1^{+2.7}_{-3.1}$ degrees.
The galaxy's orientation into the plane of the paper, $\Phi = 85.2^{+4.4}_{-3.7}$ degrees, is small.
Fiducially, inflowing material is expected to be along the galaxy major axis while outflowing material is expected to be along the galaxy minor axis.
Because of the inclination of the system relative to our line of sight, both inflowing and outflowing material are expected to be blueshifted.

\begin{figure} %Figure 9
\includegraphics[angle = 00,width=\columnwidth]{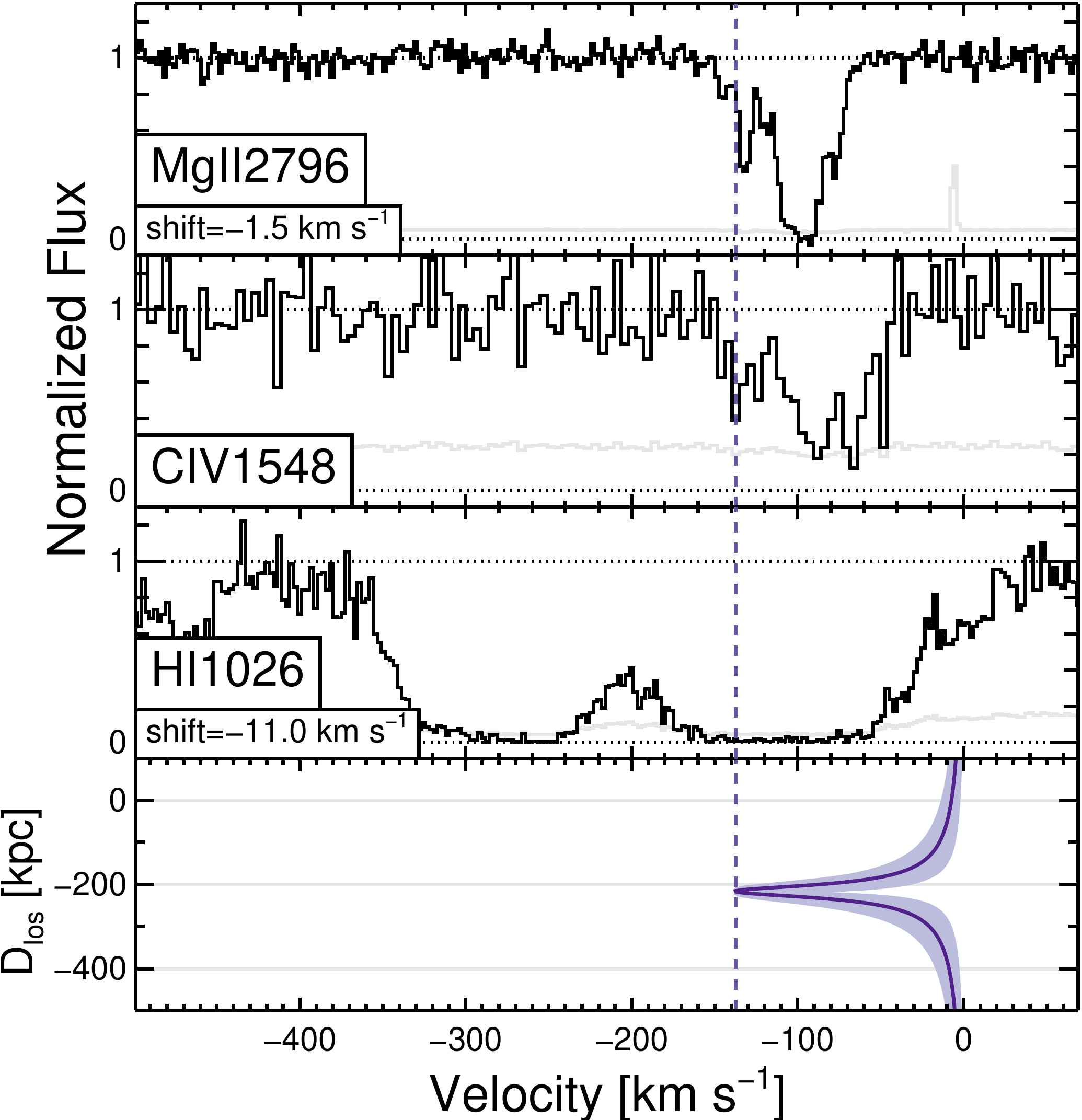}
\caption{Important absorption profiles plotted with $v_\mathrm{los}$ and $D_\mathrm{los}$ from an extended disc-like co-rotating halo model.
The light purple region corresponds to the $1\sigma$ errors in the values for $i$, $PA$, and $D$.
While most of the clouds from the low-ionization and high-ionization phases can be explained by this model, the implied $D_\mathrm{los}$ for the clouds is large, meaning that the clouds would exist far above the disc mid-plane if this model were correct.
The low-metallicity subsystem cannot be explained by this model.
$D_\mathrm{los} = -215$ kpc when $v_\mathrm{los}$ is at its most extreme value.
\label{fig:vlos}}
\end{figure}

We investigate the possibility of our clouds co-rotating within a `cold-flow disc,' in which material that is accreting onto the galaxy arises due to filamentary structure from the cosmic web to form an extended disc structure.
As done previously by \citet{kacprzak+10a}, we investigate this by employing a simple halo model from \citet{steidel+02} with the updated values for $i$ and PA.
In order to determine if all the low-, intermediate-, and high-ionization clouds can be explained by a co-rotating thick disc, we set the lagging scale-height term to $h_v = 1000$ kpc.
This term is related to the location above the mid-plane where the lagging halo velocity component begins to matter and $v_\mathrm{los}$ becomes dominated by an exponentially declining term.
Following previous studies \citep[e.g., ][]{steidel+02, kacprzak+10b, kacprzak+10a, kacprzak+11a, kacprzak+2019, french+2020, nateghi+2021}
we are forcing the disc to be a thick, nearly-rigid rotator, where material far above the disc plane is rotating along with the disc.
This places our low- and high-ionization clouds high above the galactic disc at a high angle.
The line-of-sight distance ($D_\mathrm{los}$) and line-of-sight velocity ($v_\mathrm{los}$), can be found in Figure \ref{fig:vlos}.
The figure shows that much of the low-ionization, intermediate-ionization, and high-ionization clouds (except Cloud \#1) can be explained by a co-rotating thick disc, but the low-metallicity subsystem cannot be.
Using such a large value for $h_v$ may be extreme, but it clearly illustrates the contrast between the low-metallicity subsystem and the other gas phases.
In short, it is not at all possible that the low-metallicity subsystem is due to a cold-flow disc.
Using a more plausible (smaller) value for $h_\nu$ will not affect the disparity between the low-metallicity subsystem and the other gas phases.
We find that even by adopting a non-physically motivated scale-height of $h_v = 10$ kpc \citep[adopted from][]{ho+17}, we can still explain the metal-enriched clouds with a co-rotating thick disc (except Cloud \#1 and \#2), but we are still unable to explain the low-metallicity subsystem.
This agrees with recent simulations in which high angular momentum gas accretes via a cold-flow disc, creating large co-rotating gaseous structures in the galaxy halo \citep{stewart+13,stewart+17}.
The Q0454-220 system also agrees with observations of galaxy/absorber pairs, which have been found to exhibit disc-like or accretion kinematics over a large range of impact parameters \citep[e.g.][]{ho+17, bouche+16,burchett+13,rahmani+18}.

In contrast to the low- and high-ionization material, the low-metallicity subsystem is more difficult to explain.
The subsystem exists at a much higher velocity relative to G1 and cannot be kinematically aligned with a cold-flow thick disc.
The subsystem is moving in the same direction as the low- and high-ionization material, but it has much higher angular momentum than the low- and high-ionization phases.
The subsystem's low metallicity indicates the material may be close to pristine, so its high angular momentum may agree with LCDM predictions from simulations, where newly accreted material is expected to have greater specific angular momentum than the corresponding halo \citep{stewart+13}.
Furthermore, the metallicity of subsystem indicates that the material is infalling rather than outflowing, since simulated LLSs ($16.2 < \log N_\mathrm{HI} < 19.0$) via the FIRE\footnote{http://fire.northwestern.edu/} project that have low metallicity below $\XH < -2$ tend to have infalling radial velocities, indicating that most (but not all) low metallicity LLSs are likely accreting gas \citep{hafen+17,hafen+19}.

\citet{kacprzak+11b} find a strong correlation between $W_r(2796)$ and $i/D$, which supports models made by \citet{stewart+11} that indicate an extended coldflow disc of accreting gas co-rotates with the central disc.
However, for the Q0454-220 system, not all of the low-ionization phase can be explained with a coldflow disc.

The high metallicity of the low- and high-ionization phases indicates that the gas must have been enriched by supernovae.
Cosmological simulations suggest that infalling gas at $z < 1$ is primarily composed of material that was ejected in a wind \citep{ford+14,oppenheimer+10}.
 \citeauthor{ford+14} find that at an impact parameter of $\sim 100$ kpc, most \ion{Mg}{ii} absorption (low-ionization) arises from recycled accretion, in which the same material has been both accreted and outflown in cycles, while most \ion{O}{vi} and \ion{C}{iv} absorption (high-ionization) arises almost exclusively from ancient outflows, in which particles have been ejected more than 1 Gyr previously and were not re-accreting.
 Using this as a model for the Q0454-220 absorption system, the low-ionization clouds are likely from recycled accretion and the high-ionization clouds are likely from ancient outflows.
 However, it is important to be careful with this approach, since this simply indicates the \textit{most likely} origin for the clouds in a broad sense, and does not specify how the gas is kinematically accreted or expelled.
The FIRE simulations indicate that metal-rich gas can arise either in IGM accretion or in winds, and so metallicity alone is not a completely reliable diagnostic of inflows and outflows based on kinematics relative to central galaxies \citep{hafen+17,hafen+19}.

It is useful to consider where the clouds are physically located relative to the galaxy.
The low-metallicity subsystem probably does not exist along the galaxy's poles, since outflowing material would enrich the metallicity content of the clouds.
We cannot definitively say where the low-ionization and high-ionization phases reside.
The physical sizes of the low-ionization clouds differ greatly from the sizes of the high-ionization clouds: the low-ionization clouds have thicknesses of 10s to 100s of parsecs, the high-ionization clouds have thicknesses of several kiloparsecs, and the low-metallicity subsystem clouds have completely indeterminate sizes.
Because the high-ionization and low-ionization phases are closely aligned kinematically, it is tempting to think of the higher density low-ionization clouds as being embedded within the lower density high-ionization clouds.

\subsection{Nitrogen Deficiency} \label{discussion:nitrogen}
Our model predicts the depth of the \ion{N}{v} lines in the high-ionization phase, but it underpredicts the depth of the \ion{O}{vi} lines, indicating a deficiency in nitrogen relative to oxygen and carbon relative to the solar abundance pattern.
This is not unusual for intervening QALs systems, particularly LLSs.
The multiple cloud weak \ion{Mg}{ii} system at $z=1.32$ toward PG 0117+213 has three high-ionization clouds detected in \ion{C}{iv} and \ion{O}{vi}, which have nitrogen deficiencies (relative to oxygen and carbon) of 0.5-1.2 dex \citep{masiero+05}.
The metallicities of those clouds were inferred to be $\FeH \sim -0.3$.

Another example of a nitrogen deficient abundance pattern (by $\sim$0.8 dex) is found in the offset, broad high-ionization components in the $z=1.04$ system toward PG 1634+706 \citep{zonak+04}.
In this system, the high-ionization components have inferred metallicities higher than the low-ionization components.
\citeauthor{zonak+04} hypothesize that the diffuse, high-ionization gas might be relatively undiluted supernova ejecta.

Although there are many possible production channels, nitrogen is generally believed to arise primarily from intermediate mass stars, where it is synthesized from oxygen and carbon as part of the CNO cycle in the hydrogen-burning shell \citep{pettini+02}.
This means nitrogen production lags behind oxygen, which is produced in Type II supernovae.
A low nitrogen abundance relative to oxygen can thus arise from a stellar population with recent star formation.
Low $N_\mathrm{N} / N_\mathrm{O}$ is found in low metallicity DLAs and in young dwarf galaxies in the local universe \citep{pilyugin+03}.

The metallicities of the high-ionization clouds of the Q0454-220 system are poorly constrained, so we can only claim that it is plausible that the gas is either related to a dwarf galaxy or to material related to recent star formation.

\subsection{Low-Metallicity Subsystem} \label{discussion:lowZ}
The Q0454-220 system was chosen for investigation due to its \ion{Mg}{ii} lines, which should impose a bias towards higher metallicity clouds.
However, alongside the metal enriched low-ionization and high-ionization phases exists a component that is unique because it shows no signs of metal absorbers despite having clouds with substantial hydrogen column densities ($N_\mathrm{HI} = 18.0, 15.3$).
We attempt to find systems of similar metallicity within the literature, and find few comparable results.

\citet{lehner+13} compared 29 LLSs ($16.3 \le \log N_\mathrm{HI} < 19.0$), 29 SLLSs ($19.0 \le \log N_\mathrm{HI} < 20.3 $), and 26 DLAs ($20.3 \le \log N_\mathrm{HI}$) with $z \lesssim 1$.
\citeauthor{lehner+13} find that the metallicity distribution of the LLSs is `strikingly' bimodal, with the upper half being centred on $\langle \XH \rangle = -0.33 \pm 0.33$ and the lower half being centred on $\langle \XH \rangle = -1.57 \pm 0.24$.
This is much higher than the value for our system ($\FeH < -2.5$), and the absolute lowest metallicity in their entire sample has an upper limit of $\FeH < -2.0$.
Similarly, of the 44 systems at $z \sim 0.2$ analyzed by \citet{werk+14}, the lowest metallicity example is $\log Z = -2.2$, with most of the others being well above this value.

At $z < 1$, there are only a few other known systems with such a paucity of metals:
J1500+4836 with $z_\mathrm{abs} = 0.898$ and $\mathrm{[Mg/H]} < -2.48$ \citep{wotta+16};
possibly J1139-1350 with $z_\mathrm{abs} = 0.319$ and either $\mathrm{[Si/H]} = -2.59^{+0.58}_{-0.04}$ or $\mathrm{[Si/H]} = -1.91^{+0.20}_{-0.13}$ depending on the EBR used in the modeling \citep[HM05 and HM12 respectively, ][]{pointon+19};
J135726.26+043541.3 with $z_\mathrm{abs} = 0.328637$ and $\XH = -2.58 \pm 0.09$;
J044011.90-524818.0 with $z_\mathrm{abs} = 0.865140$ and $\XH = -2.76^{+0.34}_{-0.23}$;
J152424.58+095829.7 with $z_\mathrm{abs} = 0.728885$ and $\XH = -2.92 \pm 0.05$;
and J055224.49-640210.7 with $z_\mathrm{abs} = 0.345149$ and $\XH = -2.83 \pm 0.50$ \citep{lehner+19}. 
In order to find other comparable systems to Q0454-220, we must look at much higher redshift.
\citet{fumagalli+11} observed LLSs in SDSS J113418.96+574204.6 and Q0956+122, and assuming $\log U \geq -3$ derived upper limits on metallicities of $\FeH < -4.2$ and $\FeH < -3.8$, respectively.
This is a lower metallicity than we achieve for our system, but their systems were at a much higher redshift, $z \sim 3.1 - 3.4$ where gas is generally expected to be at lower metallicities.
If we limit our analysis to $\log U \geq -3$ as well, we find that our low-metallicity subsystem has $\FeH < -4.2$.
However, in order to better compare our work with \citeauthor{fumagalli+11}, it is probably better to limit ourselves to the same $n_\mathrm{H}$ as they, since material at redshift $z \ge 3.1$ should naturally be more highly ionized than the exact same material at redshift $z \sim 0.5$.
For \citeauthor{fumagalli+11}, $\log U = -3.0$ corresponds to $n_\mathrm{H} \sim -2.0$ at $z=3.096$.
Limiting our low-metallicity subsystem to $n_\mathrm{H} \lesssim -2.0$ gives an upper limit of $\FeH < -2.9$.
In Figure \ref{fig:PI_logZ_limit} we plot the upper limits of metallicity provided by each metal.

It seems likely that the low-metallicity subsystem is due to nearly pristine accretion.
Its low-metallicity indicates that the subsystem have not been exposed to supernovae or recycled material.
The clouds are traveling at a high velocity compared to the galaxy ($\sim 275 \, \kms$) placing them beyond a potential co-rotating cold-flow disc.
It is interesting that such pristine material can occur so close to the metal enriched low-ionization and high-ionization phases.

The Q0454-220 system is similar to six weak \ion{Mg}{ii} systems discussed by \citet{muzahid+18}, in that additional \ion{H}{i} without associated metal lines are detected within $500 \, \kms$ of the \ion{Mg}{ii} absorption.
\citeauthor{muzahid+18} suggest that this complex structure indicates that these systems exist within group environments,  as do the majority of weak \ion{Mg}{ii}
absorbers in general.
The Q0454-220 system matches this suggestion, since it has three known galaxies within $D < 320 \mathrm{kpc}$ at similar redshifts (see \S \ref{sec:galaxy}).
The low-metallicity subsystem agrees with recent data from the Multi-Unit Spectroscopic Explorer (MUSE) that indicates infalling/outflowing gas cannot necessarily be understood via a single galaxy's geometry alone and may be related to larger scale intra-group gas or structure \citep{peroux+17,bielby+17}.
However, without deeper investigation of the surrounding region (for example, via integral field spectroscopy), it is not possible to definitively conclude the nature of intra-group kinematics for the Q0454-220 system.

\section{Summary and Conclusions} \label{sec:conclusions}
We have presented a model for an absorption system along the line of sight to QSO 0454-220.
The absorption system is best modeled as ten clouds; six clouds due to gas in a low-ionization phase, two clouds due to gas in a high-ionization phase, and two clouds due to a low-metallicity subsystem with indeterminate ionization parameter that lies $\sim 200 \, \kms$ blueward of the low- and high- ionization phases.
The low-ionization material has with $\FeH \sim -0.5$ and $n_\mathrm{H} \sim 10^{-2.3} \, \mathrm{cm}^{-3}$; the high-ionization material has a conservative lower limit of $\FeH > -2.3$ and $n_\mathrm{H} \sim 10^{-3.9} \, \mathrm{cm}^{-3}$; the low-metallicity subsystem has a conservative upper limit of $\FeH < -2.5$.

We have used our detailed modeling of the absorption lines and the detailed observations of the host galaxy to determine the sources of the CGM.
The low- and high-ionization phases can be explained mostly via a cold-flow disc, with the absorption system lying high above the plane of the galaxy G1.
The high metallicity of the low-ionization phase indicates that it is accreting material, most likely material that has been expelled and re-accreted multiple times.
The metallicity of the high-ionization phase is less clear, but it is likely due to ancient outflowing material.
The low-metallicity subsystem has an extremely low metallicity compared to other similar systems with similar redshifts.
The subsystem cannot be explained via a cold-flow disc, and may instead by newly accreting material moving along a halo filament with higher angular momentum that has not interacted greatly with galactic outflows.

The existence of both high-metallicity and low-metallicity absorption along the same line of sight and the existence of several nearby galaxies may indicate the presence of a group environment.
However, this conclusion is not definitive and further spectral imaging studies are required to better constrain the system.

\section*{Acknowledgements}
This project was supported by NASA through grant HST GO-12466 from the Space Telescope Science Institute, which is operated by the Association of Universities for Research in Astronomy, Inc., under NASA contract NAS5-26555.
JCC, CWC, and JMN acknowledge support by the National Science Foundation under Grant No. AST-1517816.
JMN acknowledges support in part by the MEXT/JSPS KAKENHI Grant Number 17H01111 and 19H05810.
SM acknowledges support from ERC grant 278594-GasAroundGalaxies.
GGK acknowledges the support of the Australian Research Council through the Discovery Project (DP170103470).
GGK also acknowledges supported by the Australian Research Council Centre of Excellence for All Sky Astrophysics in 3 Dimensions (ASTRO 3D),  through project number CE170100013.
This research has made use of the services of the ESO Science Archive Facility.

\section*{Data Availability}
The \textit{HST} data underlying this article are available at the Mikulski Archive for Space Telescopes.
The \textit{VLT} data underlying this article are available at the European Southern Observatory Science Archive Facility.
The \textit{Keck} data and all other data underlying this article will be shared on reasonable request to the corresponding author.

%%%%%%%%%%% BIBLIOGRAPHY

\def\aj{AJ}%
\def\actaa{Acta Astron.}%
\def\araa{ARA\&A}%
\def\apj{ApJ}%
\def\apjl{ApJ}%
\def\apjs{ApJS}%
\def\ao{Appl.~Opt.}%
\def\apss{Ap\&SS}%
\def\aap{A\&A}%
\def\aapr{A\&A~Rev.}%
\def\aaps{A\&AS}%
\def\azh{AZh}%
\def\baas{BAAS}%
\def\bac{Bull. astr. Inst. Czechosl.}%
\def\caa{Chinese Astron. Astrophys.}%
\def\cjaa{Chinese J. Astron. Astrophys.}%
\def\icarus{Icarus}%
\def\jcap{J. Cosmology Astropart. Phys.}%
\def\jrasc{JRASC}%
\def\mnras{MNRAS}%
\def\memras{MmRAS}%
\def\na{New A}%
\def\nar{New A Rev.}%
\def\pasa{PASA}%
\def\pra{Phys.~Rev.~A}%
\def\prb{Phys.~Rev.~B}%
\def\prc{Phys.~Rev.~C}%
\def\prd{Phys.~Rev.~D}%
\def\pre{Phys.~Rev.~E}%
\def\prl{Phys.~Rev.~Lett.}%
\def\pasp{PASP}%
\def\pasj{PASJ}%
\def\qjras{QJRAS}%
\def\rmxaa{Rev. Mexicana Astron. Astrofis.}%
\def\skytel{S\&T}%
\def\solphys{Sol.~Phys.}%
\def\sovast{Soviet~Ast.}%
\def\ssr{Space~Sci.~Rev.}%
\def\zap{ZAp}%
\def\nat{Nature}%
\def\iaucirc{IAU~Circ.}%
\def\aplett{Astrophys.~Lett.}%
\def\apspr{Astrophys.~Space~Phys.~Res.}%
\def\bain{Bull.~Astron.~Inst.~Netherlands}%
\def\fcp{Fund.~Cosmic~Phys.}%
\def\gca{Geochim.~Cosmochim.~Acta}%
\def\grl{Geophys.~Res.~Lett.}%
\def\jcp{J.~Chem.~Phys.}%
\def\jgr{J.~Geophys.~Res.}%
\def\jqsrt{J.~Quant.~Spec.~Radiat.~Transf.}%
\def\memsai{Mem.~Soc.~Astron.~Italiana}%
\def\nphysa{Nucl.~Phys.~A}%
\def\physrep{Phys.~Rep.}%
\def\physscr{Phys.~Scr}%
\def\planss{Planet.~Space~Sci.}%
\def\procspie{Proc.~SPIE}%
\let\astap=\aap
\let\apjlett=\apjl
\let\apjsupp=\apjs
\let\applopt=\ao
\bibliographystyle{mnras}
\bibliography{norris_z04833_final}

% Don't change these lines
\bsp	% typesetting comment
\label{lastpage}

\end{document}